\documentclass[12pt]{iopart}

\usepackage[varg]{txfonts}
\usepackage{bm}
\usepackage{graphicx}
\usepackage{latexsym}
\usepackage{subfigure}
\usepackage{supertabular}
\usepackage{iopams}
\usepackage{hyperref}

\def\rd{\mathrm{d}}

\def\Ry{\,\mathrm{Ry}}
\def\kelvin{\,\mathrm{K}}

\begin{document}

\title[Completeness of CI and CC expansions]
      {Importance of the completeness of the configuration interaction and
       close coupling expansions in $R$-matrix calculations for highly-charged ions:
       electron-impact excitation of Fe$^{20+}$}
\author{L Fern\'{a}ndez-Menchero$^{1}$\footnote{Present address:
           Department of Physics and Astronomy, Drake University.
           Des Moines IA, 50311, USA}, A S Giunta$^{2}$,
        G Del~Zanna$^{3}$ and N~R Badnell$^{1}$}
\address{$^{1}$Department of Physics, University of Strathclyde,
           Glasgow G4 0NG, UK}
\address{$^{2}$STFC Rutherford Appleton Laboratory, Chilton, Didcot, Oxon. 
         OX11 0QX, UK}
\address{$^{3}$Department of Applied Mathematics and Theoretical Physics, 
           University of Cambridge, Cambridge CB3 0WA, UK}

\begin{abstract}

We have carried-out two intermediate coupling frame transformation (ICFT) 
$R$-matrix  calculations for the electron-impact excitation of $\mathrm{C}$-like 
$\mathrm{Fe}^{20+}$, both of which use the same expansions for their
configuration interaction (CI) and close-coupling (CC) representations.
The first expansion arises from the configurations  
$\mathrm{2s^2\,2p^2, \,2s\,2p^3, \,2p^4}$, 
$\mathrm{\{2s^2\,2p,\,2s\,2p^2,\,2p^3\}}\,nl$, 
with $n=3,4$, for $l=0 - 3$, which give rise to 564 CI/CC levels.
The second adds configurations $\mathrm{2s^2\,2p\,5l}$, for $l=0-2$,
which give rise to 590 CI/CC levels in total. 

Comparison of oscillator
strengths and effective collision strengths from these two calculations
demonstrates the lack of convergence in data for $n=4$ from the smaller one.
Comparison of results for the 564 CI/CC level calculation with an
earlier ICFT $R$-matrix calculation which used the exact same CI
expansion but truncated the CC expansion to only 200 levels
demonstrates the lack of convergence of the earlier data, particularly
for $n=3$ levels. Also, we find that the results of our 590 CC $R$-matrix calculation are
significantly and systematically larger than those of an earlier comparable 
Distorted Wave-plus-resonances calculation.

Thus, it is important still to take note of the (lack of) convergence in
both atomic structural and collisional data, even in such a highly-charged
ion as Fe$^{20+}$, and to treat resonances non-perturbatively. 
This is of particular importance for Fe ions given
their importance in the spectroscopic diagnostic modelling of 
astrophysical plasmas.

\end{abstract}

\submitto{\JPB}
\maketitle

\section{Introduction}
\label{sec:introduction}

Accurate data for Fe$^{20+}$ are required for solar physics 
applications.
The forbidden line of $\mathrm{Fe\,XXI}$ at $1354.1\,\AA$ is observed 
routinely with the Interface Region Imaging Spectrograph 
(IRIS \cite{depontieu2014}), launched in July 2013. 
IRIS has been producing excellent spectra and images of the solar atmosphere 
at very high resolution, and this line is used to study solar flares 
(see, e.g. \cite{young2015,polito2015}).

The $n=2 \to n=2$ transitions in the soft X-rays are known to be excellent
density diagnostics for solar flare plasma, as shown e.g. 
in \cite{mason1984,delzanna2013}. 
These lines have been observed routinely from 2010 until 2014 with the  
Extreme ultraviolet Variability Experiment (EVE) instrument on-board the 
Solar Dynamics Observatory (SDO).
The resonance line, at $128.7\,\AA$, during solar flares, becomes the dominant
contribution to the $131\,\AA$ band of the SDO Atmospheric Imaging Assembly 
(AIA), as described in \cite{odwyer2010,petkaki2012}.
Images from the AIA $131\,\AA$ band are routinely used since 2010 to study 
solar flares.

All of the iron ions have been studied during the course of the IRON 
project \cite{hummer1993}.
Several works exist for the electron-impact excitation of 
$\mathrm{C}$-like $\mathrm{Fe}^{20+}$ spanning several decades.
Bhatia {\em et al} \cite{bhatia1987} calculated energies and transition line 
strengths for the $\mathrm{C}$-like ions from $\mathrm{Ar}^{12+}$ up 
to $\mathrm{Kr}^{30+}$.
Zhang and Sampson \cite{zhang1996} calculated collision strengths with a
distorted wave method for all the $\mathrm{C}$-like ions from 
$\mathrm{F}^{3+}$ up to $\mathrm{Xe}^{48+}$.
They calculated all of the transitions within the $n=2$ levels.
Aggarwal \cite{aggarwal1991} used the Dirac-Fock $R$-matrix method including
in the configuration interaction (CI) and close coupling (CC) expansions all 
the configurations of $n=2$, which leads to a total of 20 fine-structure levels.
Aggarwal and Keenan \cite{aggarwal1999b} later improved upon their work by
using a more extended basis set, now including some configurations of $n=3$, 
for a total of 46 levels.

The previous work of Badnell and Griffin~\cite{badnell2001a} used the
Intermediate Coupling Frame Transformation (ICFT) $R$-matrix 
method \cite{griffin1998}.
The atomic structure was quite accurate, including in the CI expansion
all of the configurations up to $n=4$, obtaining a total of 564 levels.
In the CC calculation only 200 levels were included of these 564 from the
CI calculation. However, these 200 levels did not correspond to the
lowest energetic ones. The goal of that work was to look at the effect
of resonances attached to $n=4$ levels on transitions from the ground
configuration. 
To that end, the CC expansion included all $\mathrm{2s^2\,2p}\,4l$
configurations but omitted the $\mathrm{2p^3}\,3l$ 
(as well as all other $n=4$ ones.)
The argument being that the omitted ones (especially $n=3$) could only
couple weakly with the ground configuration. Computationally, it is
now possible to investigate this.

The most recent work on $\mathrm{Fe}^{20+}$ appears to be that of 
Landi and Gu \cite{landi2006b}\footnote{
Online material: \url{http://iopscience.iop.org/article/10.1086/500286/fulltext/tables.tar.gz}}
 who used a Distorted Wave (DW) method which 
took account of resonances attached to the $n=2$ levels and the most
important $n=3$ levels, in the independent processes and
isolated resonance approximations.
Their CI expansion was a little larger, including three configurations 
with $n=5$, for a total of 590 levels.
Landi and Gu  calculated collision strengths and effective collision strengths 
for electron-impact excitation from the ground and first two excited levels 
of $\mathrm{Fe}^{20+}$.

In present work we first include the 564 levels of Badnell and Griffin~\cite{badnell2001a} 
in the CI \& CC calculations. We use exactly the same atomic structure as in 
\cite{badnell2001a} so that we can study purely the completeness
of the CC expansion.
The effect of the completeness of the CI and CC expansions was studied in 
detail in \cite{fernandez-menchero2015b}.
In consequence we expect changes in the results of the effective collision
strengths, mostly for transitions to the levels close to the cut-off of
the basis set in \cite{badnell2001a}.
The main effect should be due to additional resonance enhancement, which 
affects most strongly the weakest (forbidden) transitions at low 
temperatures ($<10^{6}\kelvin$).

In addition, we extend our calculation to use the same
configuration expansion as Landi and Gu \cite{landi2006b}, by including the
$\mathrm{2s^2\,2p}\,5l$ with $l=0-2$, for a total of 590 CI \& CC levels.
The addition of those 26 levels will improve the convergence of the expansion,
mostly for the levels $\mathrm{2s^2\,2p}\,4l$ with $l=0-2$.

Finally, $\mathrm{Fe}^{20+}$ DW data were calculated with the {\sc autostructure} 
program \cite{badnell2011b}  as part of the baseline data improvement initiative for fusion 
within the EU FP7  programme\footnote{\url{http://adas-eu.ac.uk}}
and the results were made available via the OPEN-ADAS database\footnote{\url{http://open.adas.ac.uk}}.
The target CI expansion used the same set of configurations as the present one
and \cite{badnell2001a}, i.e. 564 levels, but the atomic structure was not optimized.
Here we check the validity of these results by comparing them with 
the (DW) ones we obtain using the optimized target of \cite{badnell2001a}. 
We also look at the effect of unitarization on these DW results.

The paper is organised as follows. 
In section \ref{sec:structure} we give details of the description of the 
atomic structure and 
in section \ref{sec:scattering} that of our $R$-matrix and
distorted wave calculations.
In section \ref{sec:results} we show some representative results and compare 
them with the results of the $R$-matrix calculations by \cite{badnell2001a}
and with distorted wave ones. 
The main conclusions are presented in section \ref{sec:conclusions}. 

The atomic data will be made available at our APAP network web 
page\footnote{\url{http://www.apap-network.org}}. 
They  will also be uploaded online in
the CHIANTI atomic database\footnote{\url{http://www.chiantidatabase.org}}
\cite{landi2013} and the Atomic Data and Analysis
Structure one (OPEN-ADAS).

This work is part of the UK APAP Network and provides a template for
treating the C-like isoelectronic sequences to follow our previous 
work on the
$\mathrm{Be}$-like \cite{fernandez-menchero2014a},
$\mathrm{Mg}$-like \cite{fernandez-menchero2014b},
$\mathrm{F}$-like \cite{witthoeft2007}, 
$\mathrm{Ne}$-like \cite{liang2010a},
$\mathrm{Li}$-like \cite{liang2011} and 
$\mathrm{B}$-like \cite{liang2009a}. 

Atomic units are used unless otherwise specified.

\section{Structure}
\label{sec:structure}

We want to compare results for two different CC expansions.
To do so unambiguously we must keep the atomic structure exactly the same, 
as in \cite{badnell2001a}.
Badnell and Griffin~\cite{badnell2001a} used the {\sc autostructure} program \cite{badnell2011b}.
{\sc autostructure} carries-out a diagonalization of the Breit--Pauli
Hamiltonian \cite{eissner1974} to obtain the eigenstates and energies of the 
target.
Relativistic terms, viz. mass-velocity, spin-orbit, and Darwin, are included 
as a perturbation.
The multi-electron electrostatic interactions are described by a 
Thomas-Fermi-Dirac-Amaldi model potential with scaling parameters 
$\lambda_{\mathrm{nl}}$.
In Badnell and Griffin~\cite{badnell2001a} the $\lambda_{\mathrm{nl}}$ were determined through 
a variational method in which the equally-weighted sum of the energies of 
all the terms was minimized.
In the present work we repeat the structure calculations with the same scaling 
parameters, but make new comparisons. We note that a non-optimized structure
corresponds to setting all scaling parameters to unity.

The calculation of Badnell and Griffin \cite{badnell2001a} included a total 
of 10 atomic orbitals in the basis set:
$1\mathrm{s}$, $2\mathrm{s}$, $2\mathrm{p}$, 
$3\mathrm{s}$, $3\mathrm{p}$, $3\mathrm{d}$, 
$4\mathrm{s}$, $4\mathrm{p}$, $4\mathrm{d}$, $4\mathrm{f}$. 
In the configuration interaction expansion are included all the configurations 
$\{(1\mathrm{s}^2)\}$ $\mathrm{2s^2\,2p^2}$,
$\mathrm{2s\,2p^3}$, $\mathrm{2p^4}$,
$\mathrm{2s^2\,2p}\,nl$,
$\mathrm{2s\,2p^2}\,nl$,
$\mathrm{2p^3}\,nl$
for all $nl$  orbitals previously mentioned with $n=3,4$,
for a total of 24 configurations.
The configuration list detailed above gives rise to a total 
of 268 LS terms, which on recoupling to take account of the spin-orbit 
interaction, give rise to 564 levels.
We have performed a new calculation including three additional orbitals
$5\mathrm{s}$, $5\mathrm{p}$, $5\mathrm{d}$.
In addition to the configurations for the previous calculation, we included
the $\mathrm{2s^2\,2p}\,5l$ with $l=0-2$.
This new configuration list rises to a total of 282 LS terms and 590 IC 
levels.
This is the same configuration list that was used by Landi and Gu
\cite{landi2006b}, which is the most complete to-date.
The minimized values of the scaling parameters for the present and previous
work \cite{badnell2001a} are
   $\mathrm{1s}$ $1.37988$;
   $\mathrm{2s}$ $1.25035$;
   $\mathrm{2p}$ $1.18359$;
   $\mathrm{3s}$ $1.38480$;
   $\mathrm{3p}$ $1.25830$;
   $\mathrm{3d}$ $1.39690$;
   $\mathrm{4s}$ $1.32721$;
   $\mathrm{4p}$ $1.25440$;
   $\mathrm{4d}$ $1.37130$;
   $\mathrm{4f}$ $1.44540$.
For present 590-level calculation, the scaling parameters for the three
added orbitals are
   $\mathrm{5s}$ $1.76478$;
   $\mathrm{5p}$ $1.41353$;
   $\mathrm{5d}$ $1.45655$.

In table \ref{tab:energiesfe20}, we show the present intermediate coupling (IC) 
energies and compare them with the ones obtained by Badnell and Griffin 
\cite{badnell2001a}, Landi and Gu~\cite{landi2006b}, and the observed ones 
tabulated in the CHIANTI database; all for the 50 lowest target levels.
The rest of the calculated level energies can be found online.
The experimental data collected for CHIANTI contain data from the works of
\cite{feldman2000a, martin1999b, brown2002, fawcett1987, landi2005, palmeri2003}
The agreement of the present energies with the observed values is
within $1.5\%$, with a few exceptions in the lower excited singlet levels,
and the relative errors are smaller in the present work than in previous 
theoretical ones with smaller basis sets, and more or less equal to
the ones of \cite{landi2006b} with the same configuration set.
In some cases the lifetimes differ a larger quantity between the different
atomic structures. 
For example in the case of 50th level it is more than a factor 2.
This discrepancy is due to the level mixing, level 50 mixing is
$\mathrm{2s\,2p^2\,3p\,^5D_{0  }^{o}}$ ($75\%$), plus
$\mathrm{2s^2\,2p\,3d\,^3P_{0  }^{o}}$ ($17\%$), plus
$\mathrm{2s\,2p^2\,3p\,^3P_{0  }^{o}}$ ($5\%$).
There is a part of $22\%$ which connects with the ground state in an 
intense dipole transition, while the $75\%$ correspond to a forbidden
spin-change one.
A slight change in the mixing of the least weight part will produce a large
change in the life time, but not so in the energy.

\begin{table}
   \caption{$\mathrm{Fe}^{20+}$ target level energies ($\mathrm{cm}^{-1}$) and lifetimes ($\mathrm{s}$). }
   \label{tab:energiesfe20} 
\begin{center}
\begin{scriptsize}
\begin{tabular}{|rr@{\ }r@{\quad}r@{\quad}r@{\ (}r@{)\quad}r@{\ (}r@{)\quad}r@{\ (}r@{)\quad}r@{\quad}r@{\quad}r@{\ }|}
   \hline
   \hline
   $i$ & Conf. & Level & 
   $E_{\mathrm{CHIANTI}}$ & $E_{\mathrm{th}}$ & \% & $E_{\mathrm{B01}}$ & \% & $E_{\mathrm{L06}}$ & \% & $\tau_{\mathrm{th}}$ & $\tau_{\mathrm{B01}}$ & $\tau_{\mathrm{L06}}$ \\
   \hline
     1 &      $\mathrm{2s^2\,2p^2}$ & $^3\mathrm{P}_{0  }    $  & $        0$ & $        0$ & $ 0.0$ & $        0$ & $ 0.0$ & $        0$ & $ 0.0$ & $       -    $ & $        -   $ & $        -  $   \\
     2 &      $\mathrm{2s^2\,2p^2}$ & $^3\mathrm{P}_{1  }    $  & $    73851$ & $    72710$ & $ 1.5$ & $    72595$ & $ 1.7$ & $    73041$ & $ 1.1$ & $ 1.60\,[-4] $ & $ 1.61\,[-4] $ & $1.53\,[-4] $   \\
     3 &      $\mathrm{2s^2\,2p^2}$ & $^3\mathrm{P}_{2  }    $  & $   117367$ & $   119427$ & $ 1.8$ & $   119252$ & $ 1.6$ & $   117147$ & $ 0.2$ & $ 9.41\,[-4] $ & $ 9.45\,[-4] $ & $1.17\,[-3] $   \\
     4 &      $\mathrm{2s^2\,2p^2}$ & $^1\mathrm{D}_{2  }    $  & $   244568$ & $   246885$ & $ 0.9$ & $   246568$ & $ 0.8$ & $   245710$ & $ 0.5$ & $ 3.26\,[-5] $ & $ 3.28\,[-5] $ & $3.33\,[-5] $   \\
     5 &      $\mathrm{2s^2\,2p^2}$ & $^1\mathrm{S}_{0  }    $  & $   371744$ & $   372374$ & $ 0.2$ & $   371983$ & $ 0.1$ & $   373060$ & $ 0.4$ & $ 7.39\,[-6] $ & $ 7.42\,[-6] $ & $7.50\,[-6] $   \\
     6 &        $\mathrm{2s\,2p^3}$ & $^5\mathrm{S}_{2  }^{o}$  & $   486991$ & $   478047$ & $ 1.8$ & $   477618$ & $ 1.9$ & $   479659$ & $ 1.5$ & $ 1.56\,[-8] $ & $ 1.56\,[-8] $ & $1.49\,[-8] $   \\
     7 &        $\mathrm{2s\,2p^3}$ & $^3\mathrm{D}_{1  }^{o}$  & $   776685$ & $   776317$ & $ 0.0$ & $   776093$ & $ 0.1$ & $   779724$ & $ 0.4$ & $ 7.81\,[-11]$ & $ 7.80\,[-11]$ & $7.83\,[-11]$   \\
     8 &        $\mathrm{2s\,2p^3}$ & $^3\mathrm{D}_{2  }^{o}$  & $   777367$ & $   777486$ & $ 0.0$ & $   777258$ & $ 0.0$ & $   779963$ & $ 0.3$ & $ 1.05\,[-10]$ & $ 1.04\,[-10]$ & $1.06\,[-10]$   \\
     9 &        $\mathrm{2s\,2p^3}$ & $^3\mathrm{D}_{3  }^{o}$  & $   803553$ & $   806858$ & $ 0.4$ & $   806644$ & $ 0.4$ & $   805768$ & $ 0.3$ & $ 1.37\,[-10]$ & $ 1.36\,[-10]$ & $1.39\,[-10]$   \\
    10 &        $\mathrm{2s\,2p^3}$ & $^3\mathrm{P}_{0  }^{o}$  & $   916333$ & $   915712$ & $ 0.1$ & $   915430$ & $ 0.1$ & $   920272$ & $ 0.4$ & $ 4.43\,[-11]$ & $ 4.42\,[-11]$ & $4.44\,[-11]$   \\
    11 &        $\mathrm{2s\,2p^3}$ & $^3\mathrm{P}_{1  }^{o}$  & $   924920$ & $   925643$ & $ 0.1$ & $   925364$ & $ 0.0$ & $   928822$ & $ 0.4$ & $ 4.17\,[-11]$ & $ 4.17\,[-11]$ & $4.21\,[-11]$   \\
    12 &        $\mathrm{2s\,2p^3}$ & $^3\mathrm{P}_{2  }^{o}$  & $   942364$ & $   944477$ & $ 0.2$ & $   944210$ & $ 0.2$ & $   946135$ & $ 0.4$ & $ 4.69\,[-11]$ & $ 4.69\,[-11]$ & $4.71\,[-11]$   \\
    13 &        $\mathrm{2s\,2p^3}$ & $^3\mathrm{S}_{1  }^{o}$  & $  1095679$ & $  1101047$ & $ 0.5$ & $  1100644$ & $ 0.5$ & $  1105579$ & $ 0.9$ & $ 1.00\,[-11]$ & $ 1.00\,[-11]$ & $1.02\,[-11]$   \\
    14 &        $\mathrm{2s\,2p^3}$ & $^1\mathrm{D}_{2  }^{o}$  & $  1127250$ & $  1135203$ & $ 0.7$ & $  1134994$ & $ 0.7$ & $  1137533$ & $ 0.9$ & $ 1.81\,[-11]$ & $ 1.81\,[-11]$ & $1.85\,[-11]$   \\
    15 &        $\mathrm{2s\,2p^3}$ & $^1\mathrm{P}_{1  }^{o}$  & $  1260902$ & $  1268682$ & $ 0.6$ & $  1268380$ & $ 0.6$ & $  1272627$ & $ 0.9$ & $ 1.09\,[-11]$ & $ 1.09\,[-11]$ & $1.12\,[-11]$   \\
    16 &            $\mathrm{2p^4}$ & $^3\mathrm{P}_{2  }    $  & $  1646409$ & $  1652564$ & $ 0.4$ & $  1652131$ & $ 0.3$ & $  1657412$ & $ 0.7$ & $ 1.59\,[-11]$ & $ 1.59\,[-11]$ & $1.62\,[-11]$   \\
    17 &            $\mathrm{2p^4}$ & $^3\mathrm{P}_{0  }    $  & $  1735715$ & $  1740540$ & $ 0.3$ & $  1740111$ & $ 0.3$ & $  1747301$ & $ 0.7$ & $ 1.37\,[-11]$ & $ 1.37\,[-11]$ & $1.39\,[-11]$   \\
    18 &            $\mathrm{2p^4}$ & $^3\mathrm{P}_{1  }    $  & $  1740453$ & $  1744624$ & $ 0.2$ & $  1744193$ & $ 0.2$ & $  1750849$ & $ 0.6$ & $ 1.37\,[-11]$ & $ 1.37\,[-11]$ & $1.39\,[-11]$   \\
    19 &            $\mathrm{2p^4}$ & $^1\mathrm{D}_{2  }    $  & $  1817041$ & $  1828341$ & $ 0.6$ & $  1827909$ & $ 0.6$ & $  1832103$ & $ 0.8$ & $ 2.08\,[-11]$ & $ 2.08\,[-11]$ & $2.12\,[-11]$   \\
    20 &            $\mathrm{2p^4}$ & $^1\mathrm{S}_{0  }    $  & $  2048056$ & $  2060298$ & $ 0.6$ & $  2059866$ & $ 0.6$ & $  2066463$ & $ 0.9$ & $ 1.13\,[-11]$ & $ 1.13\,[-11]$ & $1.15\,[-11]$   \\
    21 &    $\mathrm{2s^2\,2p\,3s}$ & $^3\mathrm{P}_{0  }^{o}$  & $        - $ & $  7694475$ & $  - $ & $  7694446$ & $  - $ & $  7654119$ & $  - $ & $ 4.85\,[-13]$ & $ 4.92\,[-13]$ & $4.79\,[-13]$   \\
    22 &    $\mathrm{2s^2\,2p\,3s}$ & $^3\mathrm{P}_{1  }^{o}$  & $  7661883$ & $  7704253$ & $ 0.6$ & $  7704188$ & $ 0.6$ & $  7663398$ & $ 0.0$ & $ 4.16\,[-13]$ & $ 4.22\,[-13]$ & $4.06\,[-13]$   \\
    23 &    $\mathrm{2s^2\,2p\,3s}$ & $^3\mathrm{P}_{2  }^{o}$  & $        - $ & $  7805076$ & $  - $ & $  7805048$ & $  - $ & $  7770896$ & $  - $ & $ 4.61\,[-13]$ & $ 4.68\,[-13]$ & $4.46\,[-13]$   \\
    24 &    $\mathrm{2s^2\,2p\,3s}$ & $^1\mathrm{P}_{1  }^{o}$  & $        - $ & $  7831672$ & $  - $ & $  7831515$ & $  - $ & $  7796398$ & $  - $ & $ 3.05\,[-13]$ & $ 3.09\,[-13]$ & $3.00\,[-13]$   \\
    25 &    $\mathrm{2s^2\,2p\,3p}$ & $^3\mathrm{D}_{1  }    $  & $        - $ & $  7873397$ & $  - $ & $  7873586$ & $  - $ & $  7834848$ & $  - $ & $ 2.01\,[-11]$ & $ 1.91\,[-11]$ & $1.91\,[-11]$   \\
    26 &    $\mathrm{2s^2\,2p\,3p}$ & $^1\mathrm{P}_{1  }    $  & $        - $ & $  7930234$ & $  - $ & $  7930044$ & $  - $ & $  7891979$ & $  - $ & $ 6.05\,[-12]$ & $ 5.78\,[-12]$ & $5.98\,[-12]$   \\
    27 &    $\mathrm{2s^2\,2p\,3p}$ & $^3\mathrm{D}_{2  }    $  & $        - $ & $  7933613$ & $  - $ & $  7933498$ & $  - $ & $  7895497$ & $  - $ & $ 1.40\,[-11]$ & $ 1.34\,[-11]$ & $1.33\,[-11]$   \\
    28 &    $\mathrm{2s^2\,2p\,3p}$ & $^3\mathrm{P}_{0  }    $  & $  7915463$ & $  7950515$ & $ 0.4$ & $  7950219$ & $ 0.4$ & $  7909434$ & $ 0.1$ & $ 2.89\,[-12]$ & $ 2.72\,[-12]$ & $3.11\,[-12]$   \\
    29 &    $\mathrm{2s^2\,2p\,3p}$ & $^3\mathrm{P}_{1  }    $  & $        - $ & $  8010613$ & $  - $ & $  8010478$ & $  - $ & $  7977012$ & $  - $ & $ 4.17\,[-12]$ & $ 4.05\,[-12]$ & $4.48\,[-12]$   \\
    30 &    $\mathrm{2s^2\,2p\,3p}$ & $^3\mathrm{D}_{3  }    $  & $        - $ & $  8020317$ & $  - $ & $  8020172$ & $  - $ & $  7987319$ & $  - $ & $ 1.82\,[-11]$ & $ 1.74\,[-11]$ & $1.67\,[-11]$   \\
    31 &    $\mathrm{2s^2\,2p\,3p}$ & $^3\mathrm{S}_{1  }    $  & $        - $ & $  8031268$ & $  - $ & $  8031093$ & $  - $ & $  7998341$ & $  - $ & $ 4.65\,[-12]$ & $ 4.46\,[-12]$ & $4.91\,[-12]$   \\
    32 &    $\mathrm{2s^2\,2p\,3p}$ & $^3\mathrm{P}_{2  }    $  & $        - $ & $  8037310$ & $  - $ & $  8037052$ & $  - $ & $  8002052$ & $  - $ & $ 3.36\,[-12]$ & $ 3.25\,[-12]$ & $3.69\,[-12]$   \\
    33 &    $\mathrm{2s\,2p^2\,3s}$ & $^5\mathrm{P}_{1  }    $  & $        - $ & $  8101293$ & $  - $ & $  8100865$ & $  - $ & $  8070806$ & $  - $ & $ 5.47\,[-12]$ & $ 6.54\,[-13]$ & $5.52\,[-13]$   \\
    34 &    $\mathrm{2s^2\,2p\,3p}$ & $^1\mathrm{D}_{2  }    $  & $        - $ & $  8101297$ & $  - $ & $  8100884$ & $  - $ & $  8065382$ & $  - $ & $ 6.53\,[-13]$ & $ 5.32\,[-12]$ & $6.44\,[-12]$   \\
    35 &    $\mathrm{2s^2\,2p\,3d}$ & $^3\mathrm{F}_{2  }^{o}$  & $  8074160$ & $  8114457$ & $ 0.5$ & $  8114269$ & $ 0.5$ & $  8072912$ & $ 0.0$ & $ 4.21\,[-13]$ & $ 4.22\,[-13]$ & $4.10\,[-13]$   \\
    36 &    $\mathrm{2s\,2p^2\,3s}$ & $^5\mathrm{P}_{2  }    $  & $        - $ & $  8149606$ & $  - $ & $  8149172$ & $  - $ & $  8121259$ & $  - $ & $ 6.35\,[-13]$ & $ 6.35\,[-13]$ & $6.16\,[-14]$   \\
    37 &    $\mathrm{2s^2\,2p\,3d}$ & $^3\mathrm{F}_{3  }^{o}$  & $  8118008$ & $  8153237$ & $ 0.4$ & $  8152995$ & $ 0.4$ & $  8111336$ & $ 0.1$ & $ 1.75\,[-13]$ & $ 1.74\,[-13]$ & $1.59\,[-13]$   \\
    38 &    $\mathrm{2s^2\,2p\,3d}$ & $^3\mathrm{D}_{2  }^{o}$  & $  8124085$ & $  8160595$ & $ 0.4$ & $  8160231$ & $ 0.5$ & $  8118025$ & $ 0.1$ & $ 7.08\,[-14]$ & $ 7.08\,[-14]$ & $6.93\,[-13]$   \\
    39 &    $\mathrm{2s^2\,2p\,3p}$ & $^1\mathrm{S}_{0  }    $  & $  8143710$ & $  8162776$ & $ 0.2$ & $  8162508$ & $ 0.2$ & $  8126193$ & $ 0.2$ & $ 1.50\,[-12]$ & $ 1.48\,[-12]$ & $1.40\,[-12]$   \\
    40 &    $\mathrm{2s^2\,2p\,3d}$ & $^3\mathrm{D}_{1  }^{o}$  & $  8141785$ & $  8179034$ & $ 0.5$ & $  8178623$ & $ 0.5$ & $  8135992$ & $ 0.1$ & $ 4.38\,[-14]$ & $ 4.38\,[-14]$ & $4.35\,[-14]$   \\
    41 &    $\mathrm{2s\,2p^2\,3s}$ & $^5\mathrm{P}_{3  }    $  & $        - $ & $  8199750$ & $  - $ & $  8199316$ & $  - $ & $  8170876$ & $  - $ & $ 5.80\,[-13]$ & $ 5.80\,[-13]$ & $5.57\,[-13]$   \\
    42 &    $\mathrm{2s\,2p^2\,3s}$ & $^3\mathrm{P}_{0  }    $  & $  8180254$ & $  8216612$ & $ 0.4$ & $  8216280$ & $ 0.4$ & $  8179293$ & $ 0.0$ & $ 5.02\,[-13]$ & $ 5.04\,[-13]$ & $4.71\,[-13]$   \\
    43 &    $\mathrm{2s^2\,2p\,3d}$ & $^3\mathrm{F}_{4  }^{o}$  & $        - $ & $  8232560$ & $  - $ & $  8232334$ & $  - $ & $  8195771$ & $  - $ & $ 6.82\,[-10]$ & $ 6.78\,[-10]$ & $6.62\,[-10]$   \\
    44 &    $\mathrm{2s^2\,2p\,3d}$ & $^1\mathrm{D}_{2  }^{o}$  & $        - $ & $  8240685$ & $  - $ & $  8240369$ & $  - $ & $  8204330$ & $  - $ & $ 7.76\,[-14]$ & $ 7.79\,[-14]$ & $7.97\,[-14]$   \\
    45 &    $\mathrm{2s\,2p^2\,3s}$ & $^3\mathrm{P}_{1  }    $  & $        - $ & $  8252207$ & $  - $ & $  8251886$ & $  - $ & $  8217390$ & $  - $ & $ 3.92\,[-13]$ & $ 3.93\,[-13]$ & $3.66\,[-13]$   \\
    46 &    $\mathrm{2s^2\,2p\,3d}$ & $^3\mathrm{D}_{3  }^{o}$  & $  8229642$ & $  8264432$ & $ 0.4$ & $  8264055$ & $ 0.4$ & $  8227144$ & $ 0.0$ & $ 4.77\,[-14]$ & $ 4.78\,[-14]$ & $4.84\,[-14]$   \\
    47 &    $\mathrm{2s^2\,2p\,3d}$ & $^3\mathrm{P}_{0  }^{o}$  & $        - $ & $  8276802$ & $  - $ & $  8276404$ & $  - $ & $  8243034$ & $  - $ & $ 7.83\,[-14]$ & $ 7.86\,[-14]$ & $7.20\,[-14]$   \\
    48 &    $\mathrm{2s^2\,2p\,3d}$ & $^3\mathrm{P}_{1  }^{o}$  & $        - $ & $  8276891$ & $  - $ & $  8276489$ & $  - $ & $  8241557$ & $  - $ & $ 6.05\,[-14]$ & $ 6.07\,[-14]$ & $5.87\,[-14]$   \\
    49 &    $\mathrm{2s^2\,2p\,3d}$ & $^3\mathrm{P}_{2  }^{o}$  & $  8229642$ & $  8278152$ & $ 0.6$ & $  8277744$ & $ 0.6$ & $  8241437$ & $ 0.1$ & $ 5.45\,[-14]$ & $ 5.46\,[-14]$ & $5.41\,[-14]$   \\
    50 &    $\mathrm{2s\,2p^2\,3p}$ & $^5\mathrm{D}_{0  }^{o}$  & $        - $ & $  8289029$ & $  - $ & $  8288610$ & $  - $ & $  8259742$ & $  - $ & $ 6.34\,[-13]$ & $ 6.30\,[-13]$ & $1.58\,[-12]$   \\
   \hline
\end{tabular}
\end{scriptsize}
\end{center}

      $i$: level index; Conf.: configuration; 
      Level: term/level designation (largest weight); 
      $E_{\mathrm{CHIANTI}}$: observed energy from the CHIANTI database;
      $E_{\mathrm{th}}$: theoretical level energy, present work with 590-level CI expansion;
      $E_{\mathrm{B01}}$: previous work theoretical level energy \cite{badnell2001a};
      $E_{\mathrm{L06}}$: previous work theoretical level energy \cite{landi2006b};
      \%: percentage difference between theoretical and observed data;
      $\tau_{\mathrm{th}}$: lifetimes present work;
      $\tau_{\mathrm{B01}}$: lifetimes work \cite{badnell2001a};
      $\tau_{\mathrm{L06}}$: lifetimes work \cite{landi2006b}.
      $A[B]$ denotes $A \times 10^B$.
\end{table}

To check the quality of the calculated wave functions of the target we
compare the oscillator strengths ($gf$ values) for selected transitions from
the ground state in Table \ref{tab:radiative}  
for $\mathrm{Fe}^{20+}$ with data from \cite{landi2006b}. 
Data for more transitions from ground term and other transition types can
be found in the online material.
Very good agreement, within $5\%$, is found in general.
However, the $gf$ values for the
transitions $1-54$: $\mathrm{2s^2\,2p^2\,^3P_0 - 2s\,2p^2\,3p\,^3S_1^{o}}$ 
and $1-56$: $\mathrm{2s^2\,2p^2\,^3P_0 - 2s^2\,2p\,3d\,^1P_1^{o}}$ 
differ substantially from the ones of \cite{landi2006b}.
This is due to the large term mixing of the upper states:
level $54$ has a mixing of $\mathrm{2s\,2p^2\,3p\,^3S_1^{o}}\,(42\%)$, 
$\mathrm{2s\,2p^2\,3p\,^5D_1^{o}}\,(35\%)$ and 
$\mathrm{2s\,2p^2\,3p\,^5P_1^{o}}\,(14\%)$,
and $56$ has a mixing of
$\mathrm{2s^2\,2p\,3d,^1P_1^{o}}\,(77\%)$,
$\mathrm{2s^2\,2p\,3d,^3D_1^{o}}\,(8\%)$ and
$\mathrm{2s^2\,2p\,3d,^3P_1^{o}}\,(5\%)$.
Both levels mix strong dipole allowed transitions with forbidden 
spin-change ones.
The case of transition $1-56$ mixes a $77\%$ of forbidden transition with
just a $13\%$ of dipole allowed one, so its sensitivity to the level 
mixing will be quite large, and because of that the value can differ by an
order of magnitude with the one of Landi and Gu~\cite{landi2006b}.
We find that a small variation of our scaling parameters can change the 
value of $gf$ for these transitions by a factor of two.

We have made a similar comparison for infinite energy Plane-Wave Born limits,
with Badnell and Griffin \cite{badnell2001a} only since Landi and Gu \cite{landi2006b} 
do not provide them.
We find a very similar pattern of agreement (not shown) to that shown in Table \ref{tab:radiative}
for $gf$-values.

\begin{table}
\caption{Comparison of $gf$ values for selected 
   transitions of the ion $\mathrm{Fe}^{20+}$.
   $A\,[B]$ denotes $A \times 10^B$.}
\label{tab:radiative}
\centering
\begin{tabular}{r@{\,$-$\,}lrrr}
   \hline
   \hline
   \multicolumn{5}{l}{$gf$ values: dipole transitions} \\
   \multicolumn{2}{l}{Transition} & Present work & ref~\cite{badnell2001a} & ref~\cite{landi2006b} \\
   \hline
   $  1$ & $  7$ & $ 8.730\,[-2]$ & $ 8.749\,[-2]$ & $ 8.629\,[-2]$ \\
   $  1$ & $ 11$ & $ 2.319\,[-2]$ & $ 2.317\,[-2]$ & $ 2.273\,[-2]$ \\
   $  1$ & $ 13$ & $ 3.642\,[-2]$ & $ 3.638\,[-2]$ & $ 3.563\,[-2]$ \\
   $  1$ & $ 15$ & $ 7.903\,[-5]$ & $ 7.730\,[-5]$ & $ 6.596\,[-5]$ \\
   $  1$ & $ 22$ & $ 5.141\,[-2]$ & $ 5.039\,[-2]$ & $ 5.435\,[-2]$ \\
   $  1$ & $ 24$ & $ 9.731\,[-4]$ & $ 9.735\,[-4]$ & $ 1.119\,[-3]$ \\
   $  1$ & $ 40$ & $ 1.286\,[ 0]$ & $ 1.284\,[ 0]$ & $ 1.307\,[ 0]$ \\
   $  1$ & $ 48$ & $ 5.630\,[-3]$ & $ 5.654\,[-3]$ & $ 4.090\,[-3]$ \\
   $  1$ & $ 51$ & $ 2.150\,[-2]$ & $ 2.150\,[-2]$ & $ 2.665\,[-2]$ \\
   $  1$ & $ 54$ & $ 4.267\,[-2]$ & $ 4.276\,[-2]$ & $ 1.742\,[-2]$ \\
   $  1$ & $ 56$ & $ 2.689\,[-3]$ & $ 2.588\,[-3]$ & $ 3.482\,[-2]$ \\
   $  1$ & $ 58$ & $ 1.887\,[-1]$ & $ 1.886\,[-1]$ & $ 1.995\,[-1]$ \\
   $  1$ & $ 60$ & $ 1.342\,[-1]$ & $ 1.342\,[-1]$ & $ 1.173\,[-1]$ \\
   $  1$ & $ 70$ & $ 3.480\,[-2]$ & $ 3.480\,[-2]$ & $ 3.378\,[-2]$ \\
   $  1$ & $ 94$ & $ 3.168\,[-2]$ & $ 3.176\,[-2]$ & $ 2.921\,[-2]$ \\
   \hline
\end{tabular}
\end{table}

\begin{figure}
\centering
   \subfigure{
      \includegraphics[width=0.45\columnwidth,clip]{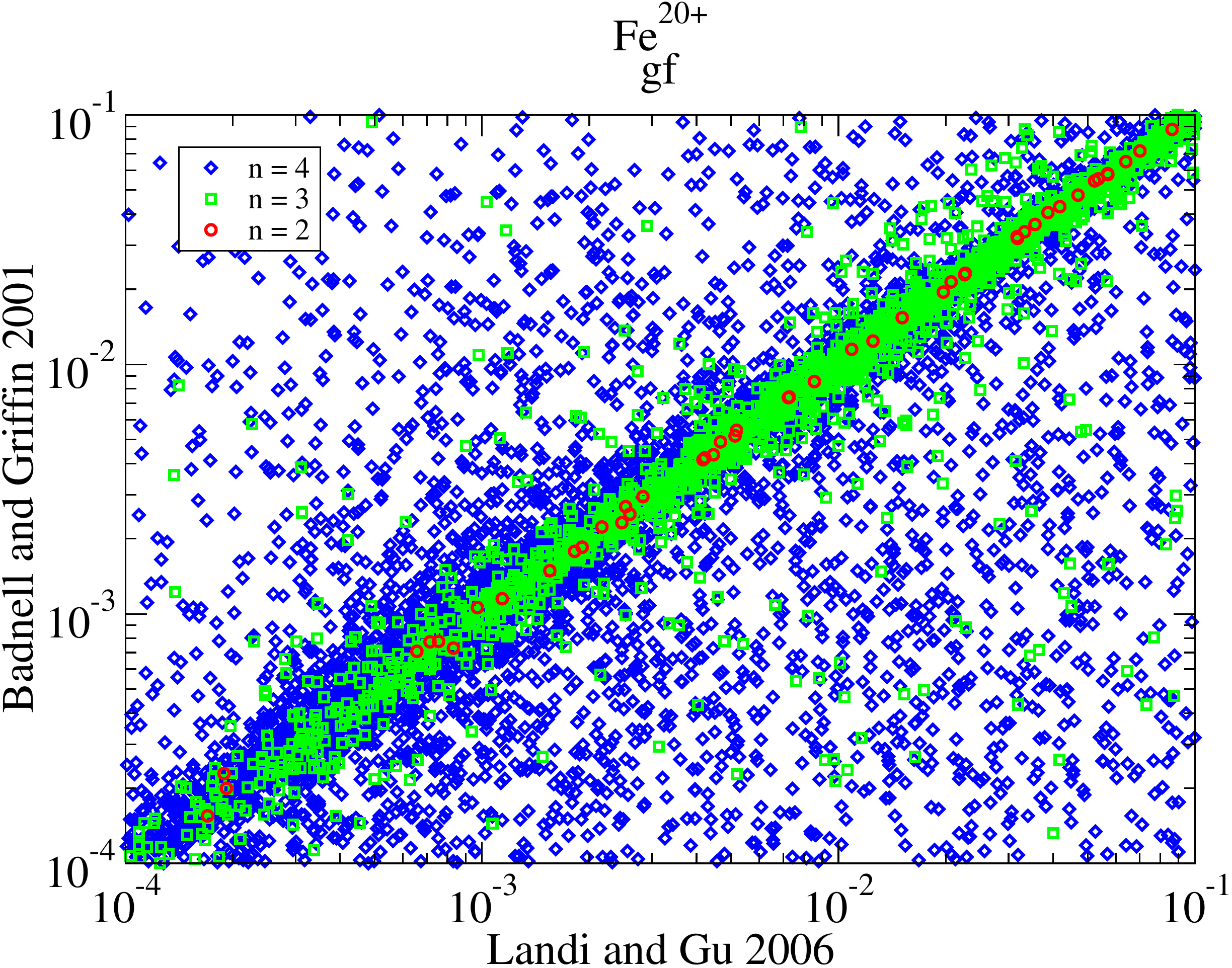}
   }\,
   \subfigure{
      \includegraphics[width=0.45\columnwidth,clip]{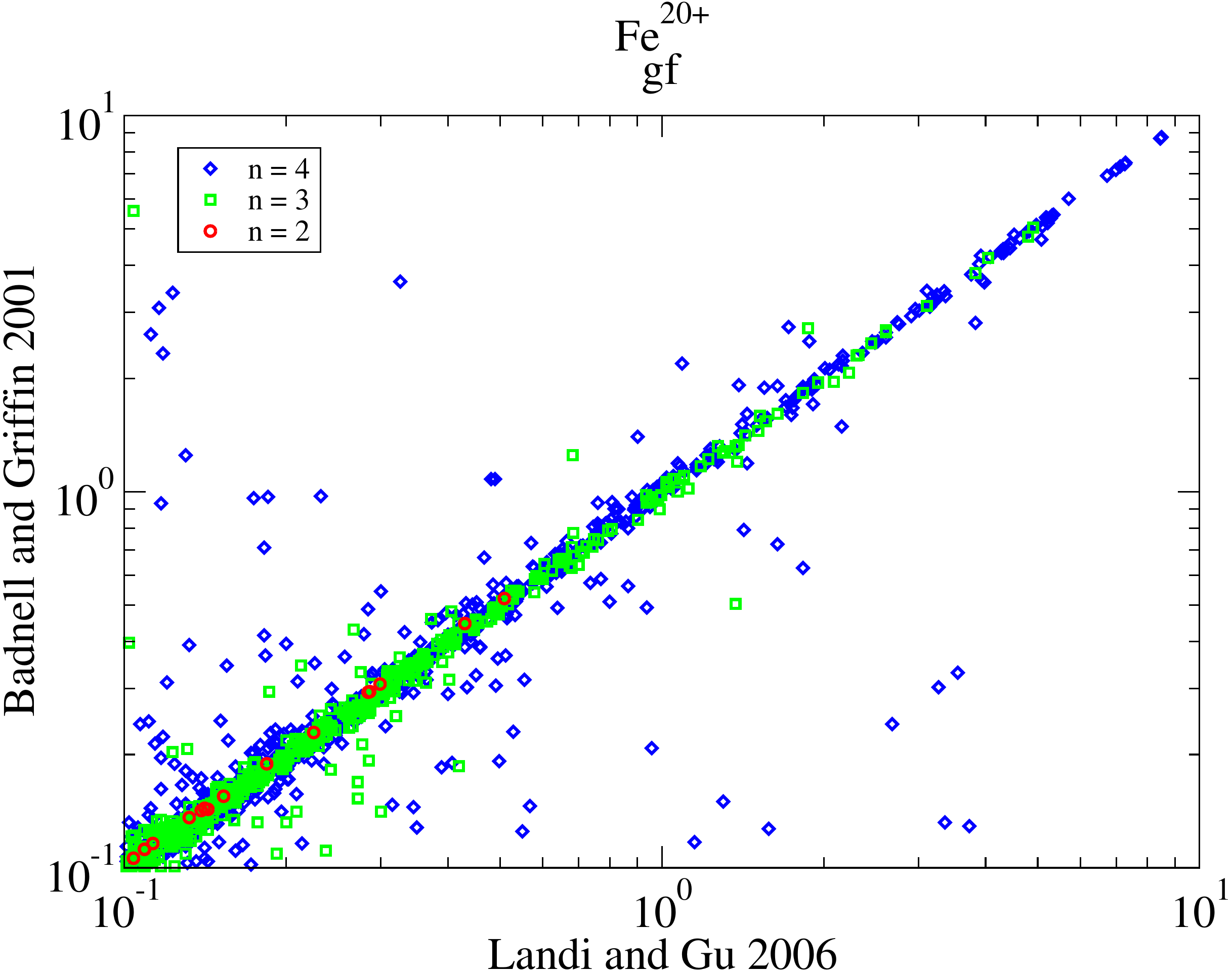}
   }
   \caption{Comparison of $gf$ for the two atomic structures of 
      works of Badnell and Griffin \cite{badnell2001a} and
      Landi and Gu \cite{landi2006b}, 564- vs 590-level CI expansions
      of $\mathrm{Fe}^{20+}$.
      $\circ$: transitions with upper level with $n=2$; 
      $\square$: transitions with upper level with $n=3$; 
      $\diamond$: transitions with upper level with $n=4$.
      Left panel: weak transitions ($gf < 0.1$);
      Right panel: strong transitions ($gf > 0.1$).
      Colour online.
      }
   \label{fig:gffe20}
\end{figure}

\begin{figure}
\centering
   \subfigure{
      \includegraphics[width=0.45\columnwidth]{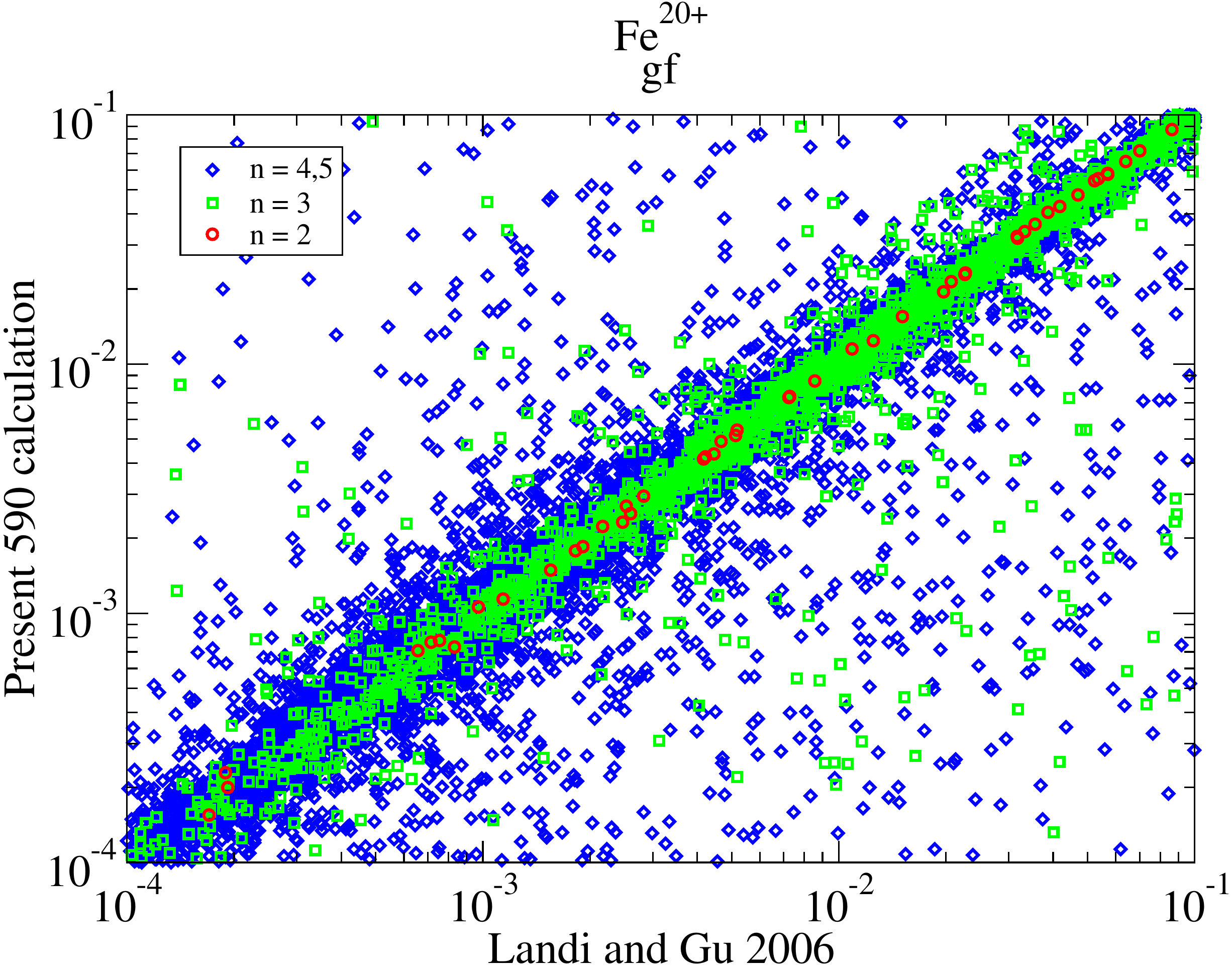}
   }\,
   \subfigure{
      \includegraphics[width=0.45\columnwidth]{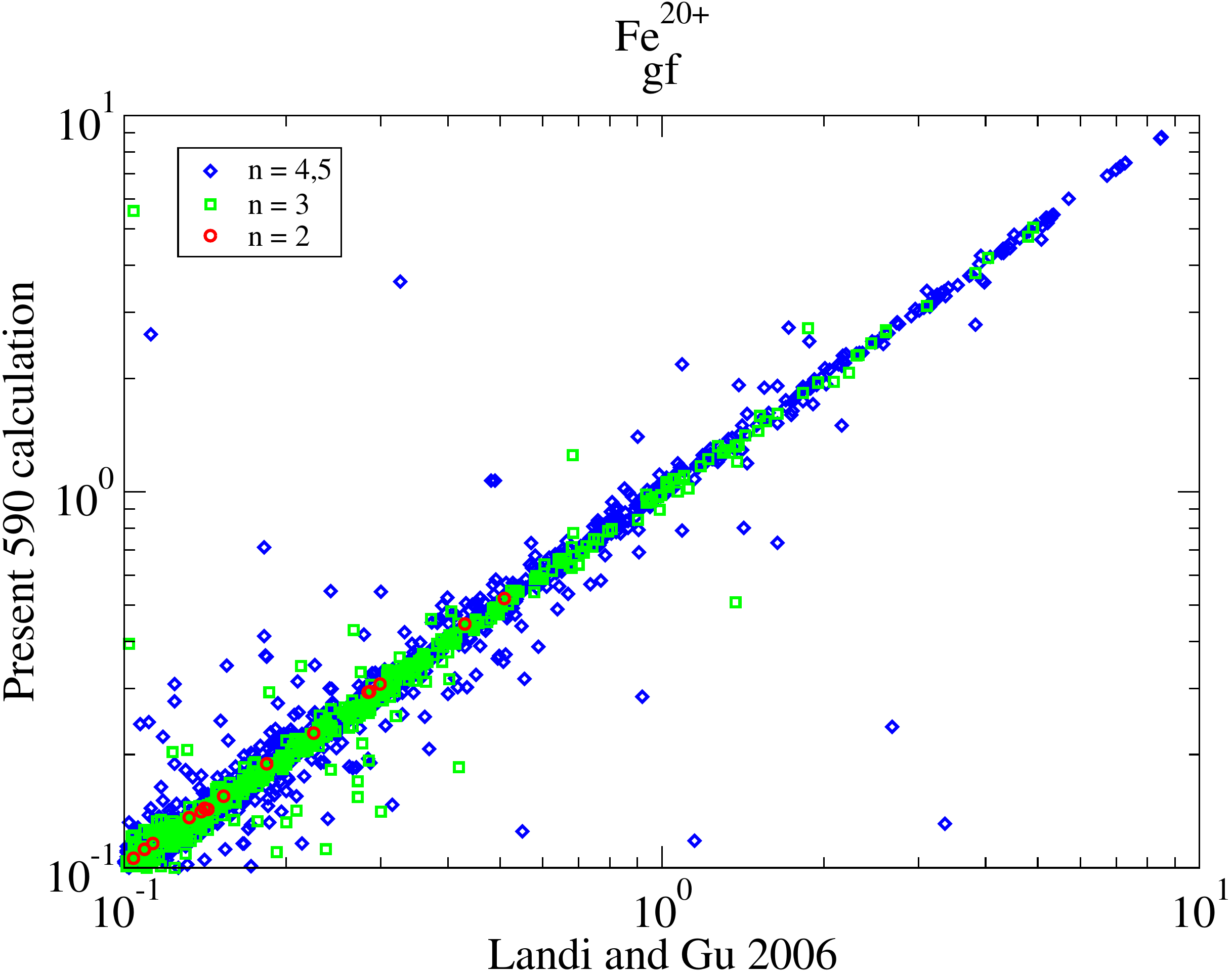}
   }
   \caption{Comparison of $gf$ for the two atomic structures of 
      works of Landi and Gu \cite{landi2006b} 590-level
      with {\sc autostructure} one with the same CI expansion
      of $\mathrm{Fe}^{20+}$.
      $\circ$: transitions with upper level with $n=2$; 
      $\square$: transitions with upper level with $n=3$; 
      $\diamond$: transitions with upper level with $n=4$.
      $\triangledown$: transitions with upper level with $n=5$.
      Left panel: weak transitions ($gf < 0.1$);
      Right panel: strong transitions ($gf > 0.1$).
      Colour online.
      }
   \label{fig:gffe20b}
\end{figure}

In fig. \ref{fig:gffe20} we show a diagram comparing the $gf$ values 
for the atomic structures of the works \cite{badnell2001a} 
and \cite{landi2006b} for the transitions between all the levels.
For transitions between levels of $n=2$ all the points spread less than 
a $10\%$ from the (diagonal) line of equality.
This demonstrates that the CI expansion is sufficiently converged for the $n=2$
states as both lead to essentially the same results.
For transitions involving states of $n=3$ the dispersion is larger.
We find some points in which both calculations differ several orders
of magnitude.
Nevertheless, most of them differ less than a factor 2.
For $n=4$ the dispersion is quite large, with many points far away of the
line of equality.
This is a consequence of the expansion not being converged for the highest excited states.
This result is to be expected, following the
work by~\cite{fernandez-menchero2015b}.
In fig. \ref{fig:gffe20b} we show the same comparison for the structure
of the present 590-level calculation with the one of \cite{landi2006b}.
This time both structures use the same CI expansion, although the orbitals themselves differ.
The dispersion is much smaller than in the previous figure.
The convergence for the levels $\mathrm{2s^2\,2p}\,4l$ with $l=0-2$
 has improved considerably.
The 26 new levels included will not increase the 
computation time by a large amount, but the quality of the atomic structure of 
the target is better now.
Levels have a large mixing between configurations of $n=4$ and $n=5$,
so the configuration can not be identified as a good quantum number
for the highest excited levels.
In table \ref{tab:densgf} we give the exact number of transitions which have
an error larger than a certain threshold in figs. \ref{fig:gffe20}
and \ref{fig:gffe20b}.

\begin{figure}
\centering
   \subfigure{
      \includegraphics[width=0.45\columnwidth]{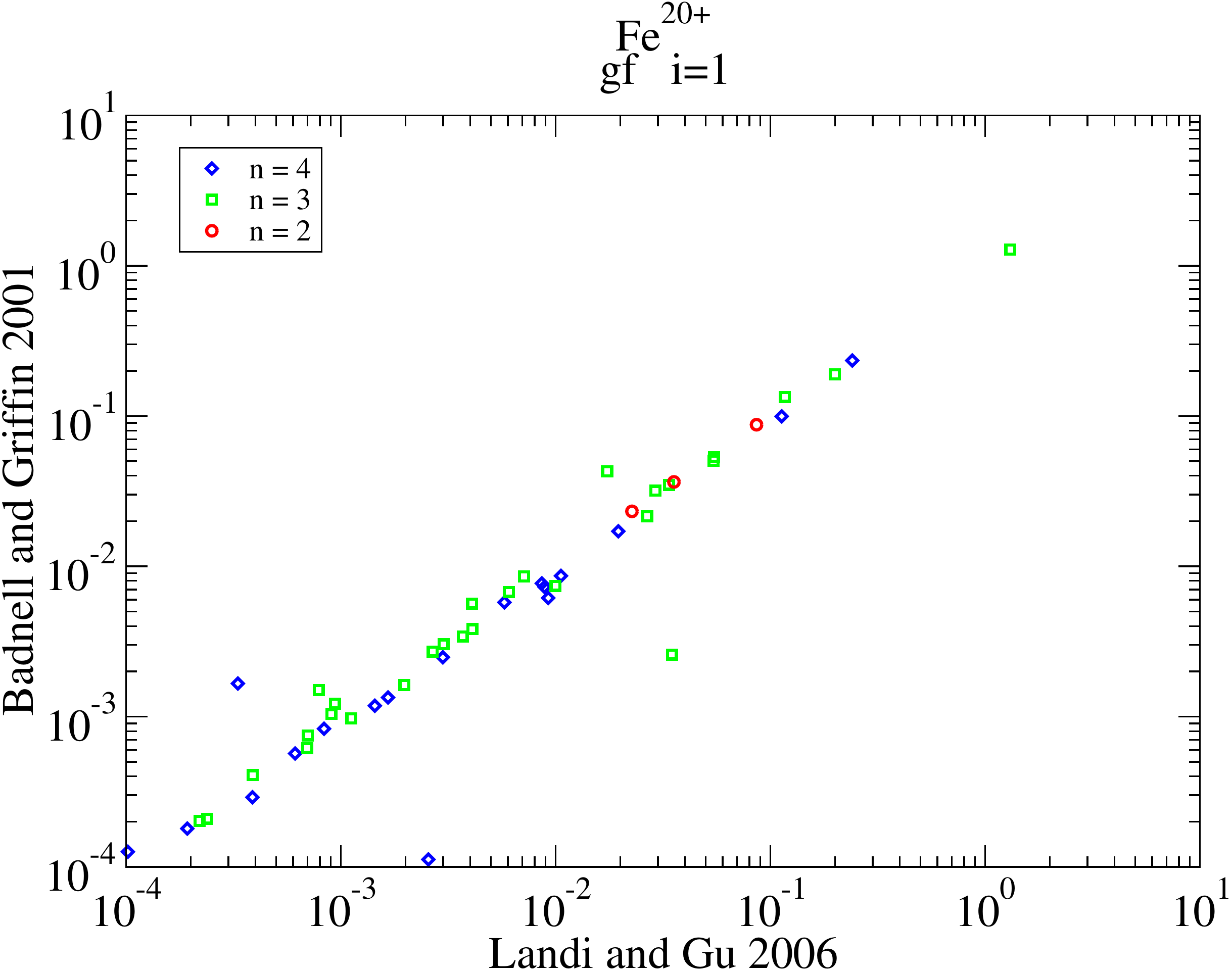}
   }\,
   \caption{Same as figure \ref{fig:gffe20}, restricting to lower
      level the ground one.
      $\circ$: transitions with upper level with $n=2$; 
      $\square$: transitions with upper level with $n=3$; 
      $\diamond$: transitions with upper level with $n=4$.
      Colour online.
      }
   \label{fig:gffe20k}
\end{figure}

In fig. \ref{fig:gffe20k} we restrict the comparison of the $gf$ factors
to the transitions from the ground level.
The levels of $n=3$ which lie far from the diagonal are states with large
term mixing between forbiden and dipole allowed transitions.
For these levels the oscillator strengths are quite sensitive to the mixing.
We see that when restricted to the ground level the data for $n=4$ are as
well converged as for $n=3$.
Transitions from the other levels of the ground term are also important for
astrophysical application and we obtain a very similar pattern of agreement (not shown)
as to the ground level.

\begin{table}
\begin{minipage}{\columnwidth}
   \caption{\label{tab:densgf} Number of transitions in Figs 
      \ref{fig:gffe20} and \ref{fig:gffe20b} which 
      differ by more than a certain relative error
      $\delta=|gf-gf_{\mathrm{LG}}|/gf_{\mathrm{LG}}$ 
      as a percentage.}
\begin{center}
\begin{tabular}{rrr}
   \hline 
     Rel. error ($\%$) & 564 \cite{badnell2001a} vs 590 \cite{landi2006b} &
     590 AS vs 590 \cite{landi2006b} \\
   \hline
     \multicolumn{3}{l}{Strong transitions ($gf > 0.1$)} \\
   \hline
     10  &   506 &   389 \\
     20  &   396 &   225 \\
     50  &   326 &   135 \\
    100  &   280 &   104 \\
    200  &   247 &    75 \\
    500  &   221 &    70 \\
   1000  &   187 &    66 \\
   \hline
     \multicolumn{3}{l}{Weak transitions ($gf < 0.1$)} \\
   \hline
     10  &  8192 &  6848 \\
     20  &  6371 &  4214 \\
     50  &  4647 &  2001 \\
    100  &   850 &   442 \\
    200  &   617 &   222 \\
    500  &   400 &   121 \\
   1000  &   267 &    86 \\
   \hline
   Total & 13816  & 14887  \\
   \hline
\end{tabular}                      
\end{center}
\end{minipage}
\end{table}

\section{Scattering}
\label{sec:scattering}

For the scattering calculation, we use the same method as in 
\cite{badnell2001a}.
It consists of an $R$-matrix formalism \cite{hummer1993,berrington1995} 
combined with an intermediate coupling frame transformation
(see \cite{badnell2001a,badnell2001b}) to include the spin-orbit mixing
efficiently and accurately.
The accuracy of the method, compared to a full Breit-Pauli $R$-matrix (BPRM)
calculation has been studied most recently in \cite{fernandez-menchero2015b}.
The differences between the ICFT and BPRM methodologies are swamped by
the uncertainties and inaccuracies due to the use of truncated CI and CC expansions.

In the $R$-matrix formalism, the configuration space is divided in two regions: 
inner and outer.
In this calculation we also split the inner region calculation in two
parts: exchange and non-exchange.
In the part including the electron exchange effects we included angular
momenta up to $2J=23$.
In the part that we neglected the exchange effects, we increased the 
maximum angular momentum to $2J=77$.
To get higher angular momenta, up to infinity, we used the top-up formula
of the Burgess sum rule \cite{burgess1974} for dipole allowed 
transitions, and a geometric series for the remaining Born allowed 
transitions (see~\cite{badnell2001a}). 
We set the number of continuum basis orbitals per angular momentum to 40, 
the smallest highest orbital energy is $1400 \Ry$.
The largest total number of channels obtained is 2870 for the calculation with 564 levels,
and 2978 for the one with 590 levels.

In the outer region, we also split the calculation in two parts.
In a low energy part, for impact energies up to the last excited level
calculated, we used a fine energy mesh step of 
approximately $3.46 \times 10^{-6} z^2 \Ry$, with $z=20$, the charge of the 
ion, to resolve the resonances sufficiently.
We extended the high energy part from the last excitation threshold to
three times the ionization potential.
In this region the collision strengths vary smoothly, so we used a coarse
mesh of $1.61 \times 10^{-4} z^2 \Ry$.
For energies above the last calculated one, we used the infinite energy limit
Plane-Wave Born $\Omega^{\infty}_{\mathrm{PWB}}$ 
and dipole line strengths $S$ from {\sc autostructure} and interpolated in a 
Burgess-Tully diagram \cite{burgess1992} for each type of transition.

To obtain the effective collision strengths $\Upsilon$ we convolute the
collision strengths $\Omega$ with a Maxwell equilibrium distribution at
an electron temperature $T$:
\begin{equation}
   \Upsilon(i-j)\ =\ \int_0^{\infty} 
   \exp \left(- \frac{E}{kT} \right)\,\Omega(i-j) \,
    \rd \left( \frac{E}{kT} \right)\,,
\label{eq:Maxwellinteg}
\end{equation}
where $E$ is the final energy of the scattered electron,
$T$ the electron temperature and $k$ the Boltzmann constant.
We calculated the effective collision strengths for electron temperatures
between $10^5$ and $10^9\,\mathrm{K}$.
That range covers the interest for both astrophysical and fusion plasmas.
Results are stored as a type 3 ADAS Atomic Data Format {\it adf04} file in 
the OPEN-ADAS database~\cite{summers1994}.

\section{Results}
\label{sec:results}

We calculated the ordinary collision strengths $\Omega$ and Maxwell-averaged 
effective collision strengths $\Upsilon$ for the electron-impact excitation 
of the ion $\mathrm{C}$-like $\mathrm{Fe}^{20+}$.
In the first calculation using the structure of \cite{badnell2001a} 
we calculated the whole transition matrix with the $R$-matrix method, 
between the $564$ fine structure levels arising in the $n=2-4$ electronic
shells, which makes for a total of $158\,766$ inelastic transitions.
For the second calculation with the larger expansion, the $590$ fine-structure 
levels give rise to a total of $173\,755$ inelastic transitions.

Firstly, we compare results of the previous work of Badnell and Griffin \cite{badnell2001a} with 
the ones we obtained with the same atomic structure but using the more complete close-coupling 
expansion.
Then we compare the DW work of Landi and Gu \cite{landi2006b} with 
our 564 CC and 590 CC level $R$-matrix calculations.
Finally, we compare a series of 564-level DW calculations with the 564 CC level $R$-matrix one,
in which we look at the effect of unitarization and the use of a non-optimized atomic structure
in the former.
The non-optimized DW work used the same CI expansion as the optimized
but we set all the scaling parameters $\lambda$ to unity.
These non-optimized non-unitarized DW results were uploaded to the OPEN-ADAS database in
2012 for use in plasma 
modelling\footnote{\url{http://open.adas.ac.uk/detail/adf04/cophps][c/dw/ic][fe20.dat}}.

\begin{figure}
\centering
   \subfigure{
      \includegraphics[width=0.45\columnwidth]{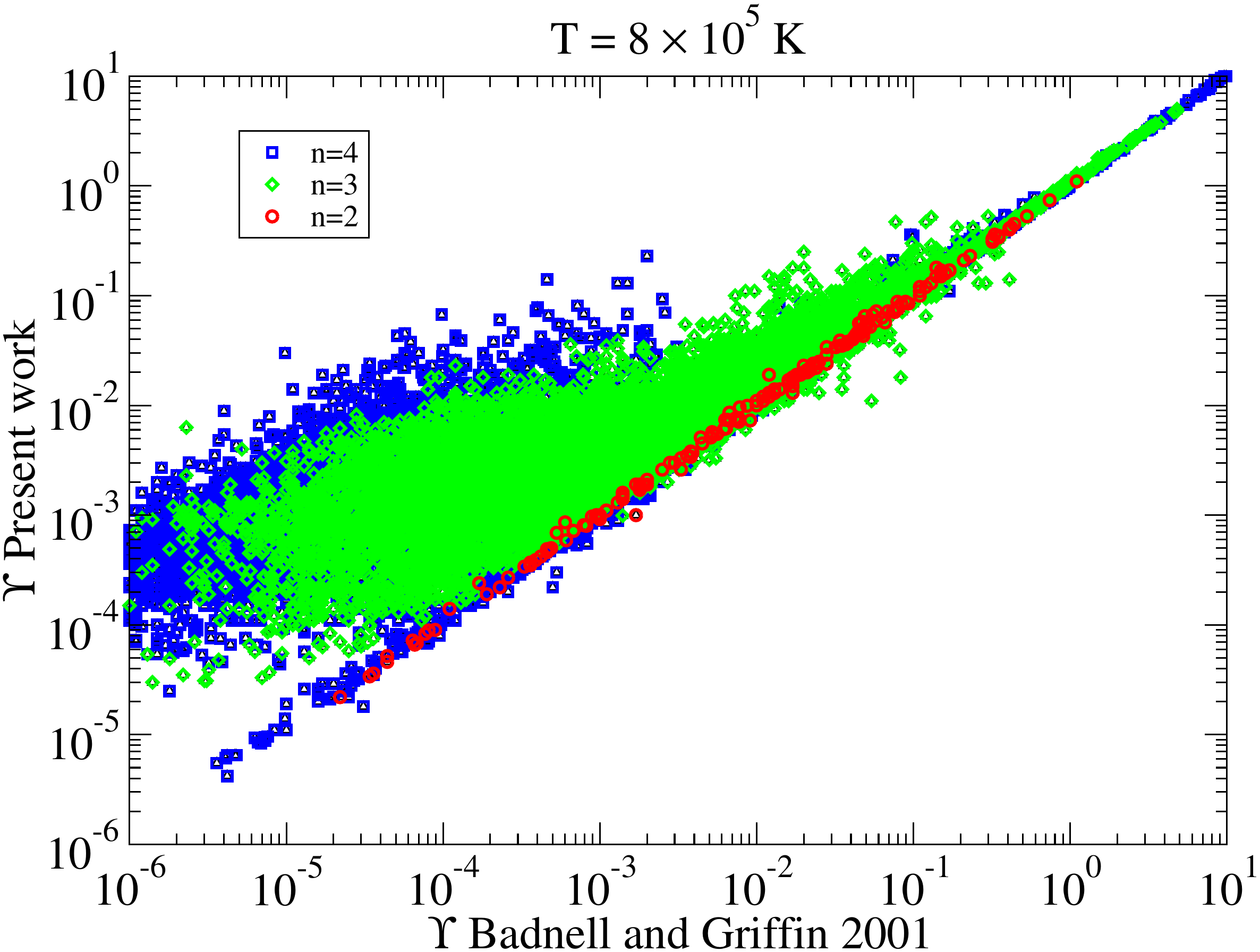}
   }\,
   \subfigure{
      \includegraphics[width=0.45\columnwidth]{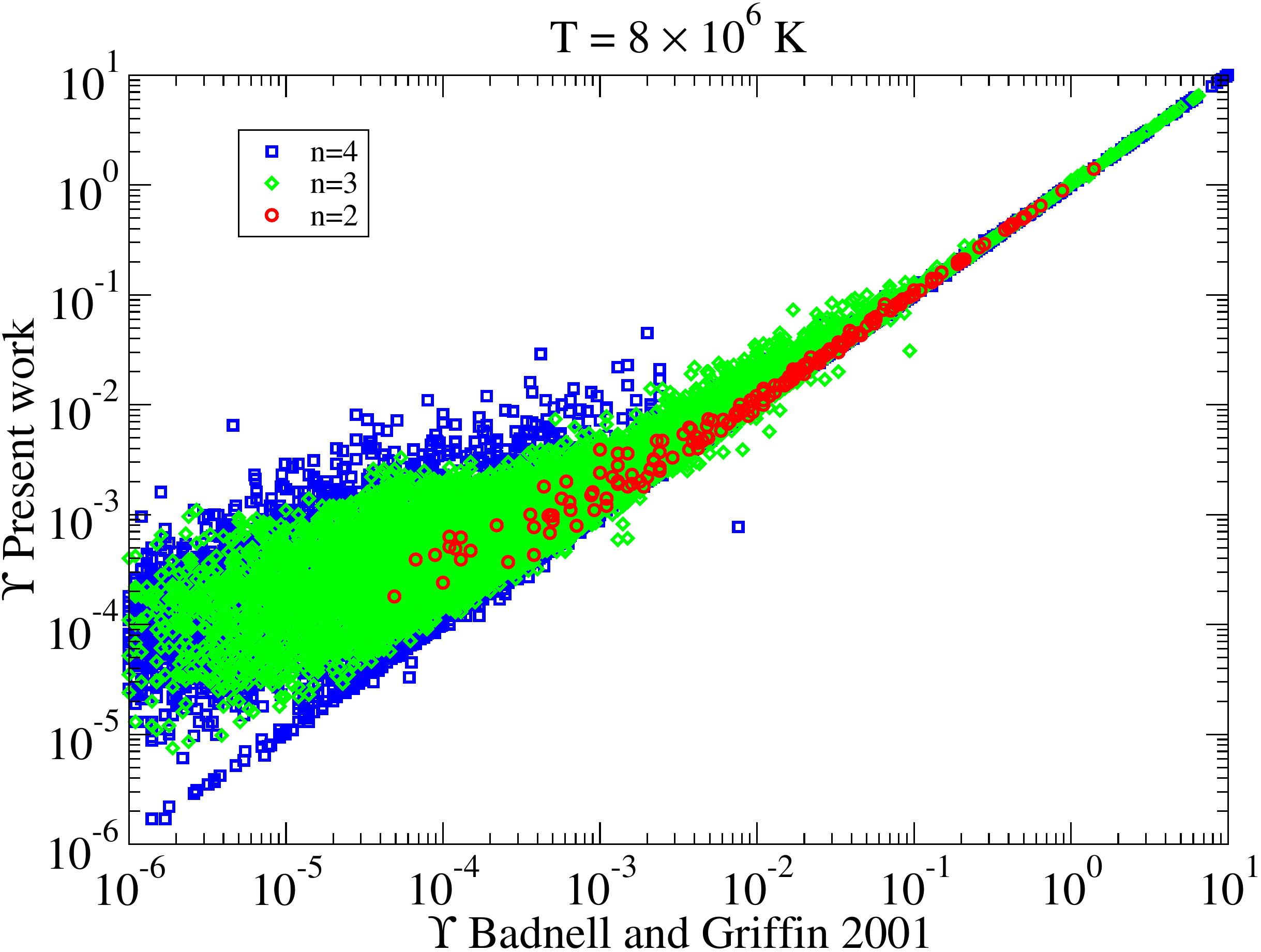}
   }\\
   \subfigure{
      \includegraphics[width=0.45\columnwidth,clip]{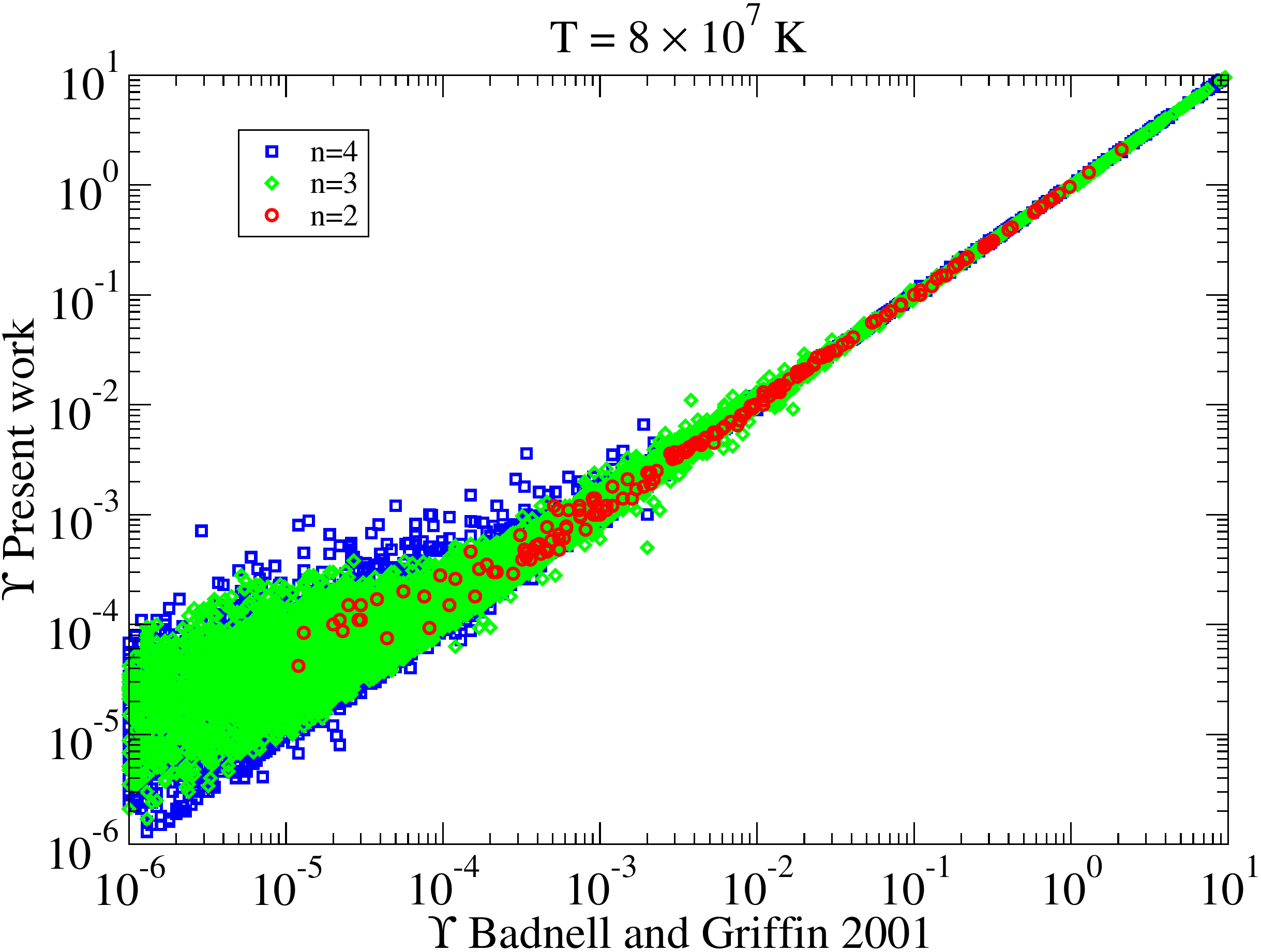}
   }
   \caption{Comparison of the $\Upsilon$ calculated with the two CC 
      expansions for all the transitions.
      $\circ$: upper level up to $n=2$.
      $\diamond$: upper level up to $n=3$.
      $\square$: upper level up to $n=4$.
      Colour online.}
   \label{fig:upscmp}
\end{figure}

Fig. \ref{fig:upscmp} shows a comparison between the $\Upsilon$ calculated
for all the transitions by Badnell and Griffin~\cite{badnell2001a} with the
ones for the present 564-level work.
Both works use exactly the same atomic structure, but different expansions
in the close coupling calculations. The present work includes all the 564
levels in the CC, while \cite{badnell2001a} only include the 200
most relevant, in  principle, to get the $\Upsilon$ for transitions  to $n=4$.
We display an intermediate temperature near to the peak abundance one, 
a lower one, and a higher one.
For weak transitions \cite{badnell2001a} underestimates the results increasingly
at low temperatures, by up to a factor of $\sim 100$ at $T=8\times10^5$~K for some of them.
This is due to the resonance enhancement associated with the additional
states in the present much larger CC expansion. The results of \cite{badnell2001a} 
show little resonant enhancement for these worst cases.
For intense transitions the resonances contribute less to the effective
collision strengths, so the underestimation disappears.
At higher temperatures the maximum of the Maxwellian distribution moves
outside the resonance region and the effect is smaller, but still large.
The number of states included in the CC calculation affects the results by a 
large amount, even when they use exactly the same atomic structure.
That is in agreement with work~\cite{fernandez-menchero2015b}.
In table \ref{tab:density} we give the exact number of values which
differ less than a certain relative error from the diagonal line of equality.

\begin{table}
\begin{minipage}{\columnwidth}
   \caption{\label{tab:density} Number of transitions in Fig 
      \ref{fig:upscmp} which 
      differ by more than a certain relative error
      $\delta=|\Upsilon_{200}-\Upsilon_{564}|/\Upsilon_{564}$ 
      as a percentage.}
\begin{center}
\begin{tabular}{rrrr}
   \hline 
     Rel. error & \multicolumn{3}{c}{Temperature ($\kelvin$)} \\
     ($\%$) & $8 \times 10^5$ & $8 \times 10^6$ & $8 \times 10^7$ \\
   \hline
     10  & 17931  & 15990  & 12375   \\
     20  & 16814  & 14775  & 10617   \\
     50  & 14872  & 12569  &  8274   \\
    100  & 13040  & 10520  &  6500   \\
    200  & 10772  &  8448  &  4946   \\
    500  &  8088  &  6186  &  3229   \\
   1000  &  6645  &  4684  &  2136   \\
   \hline
   Total & 19900  & 19900  & 19900   \\
   \hline
\end{tabular}                      
\end{center}
\end{minipage}
\end{table}

\begin{figure}
\centering
   \subfigure{
      \includegraphics[width=0.3\columnwidth]{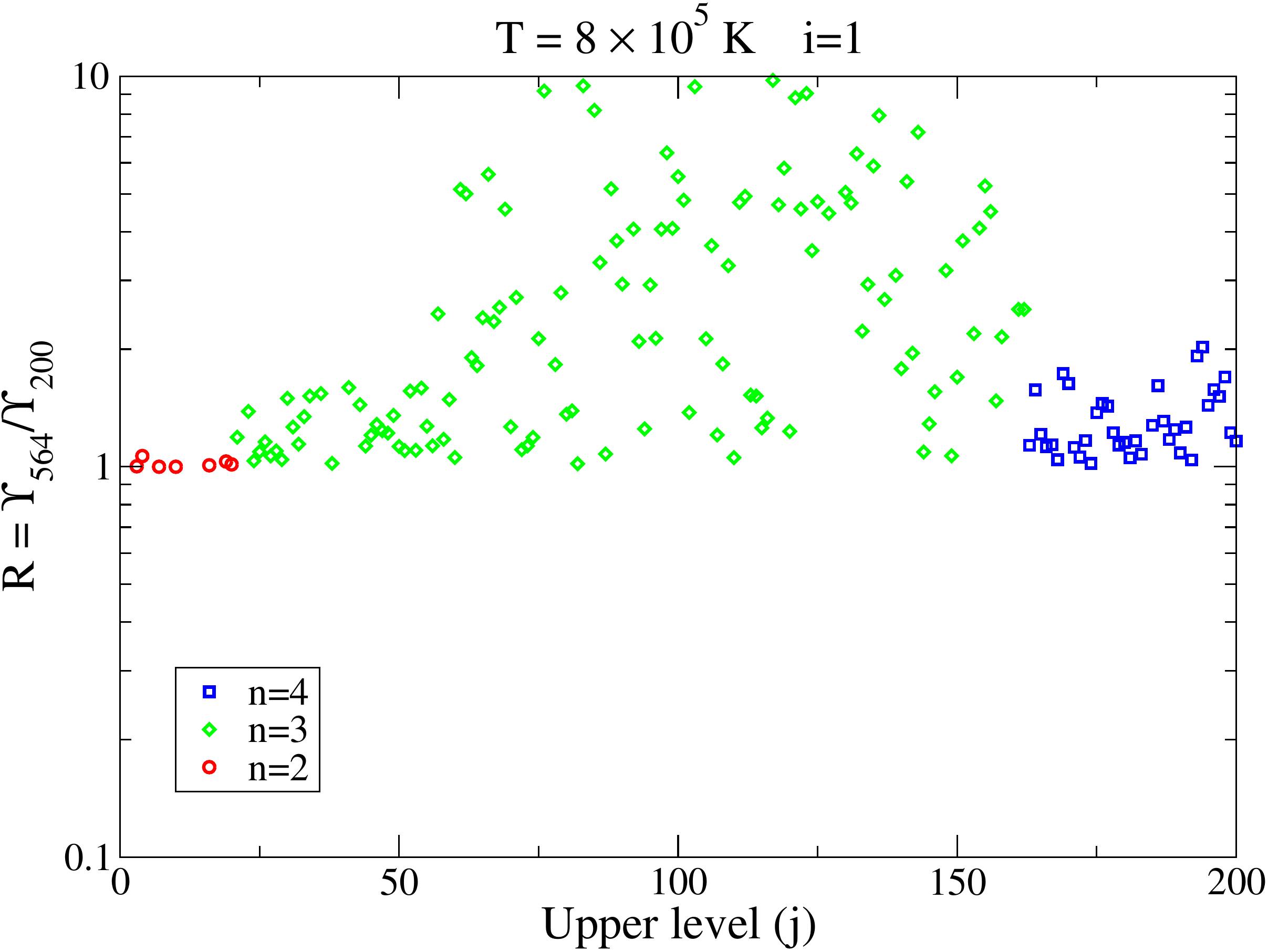}
   }\,
   \subfigure{
      \includegraphics[width=0.3\columnwidth]{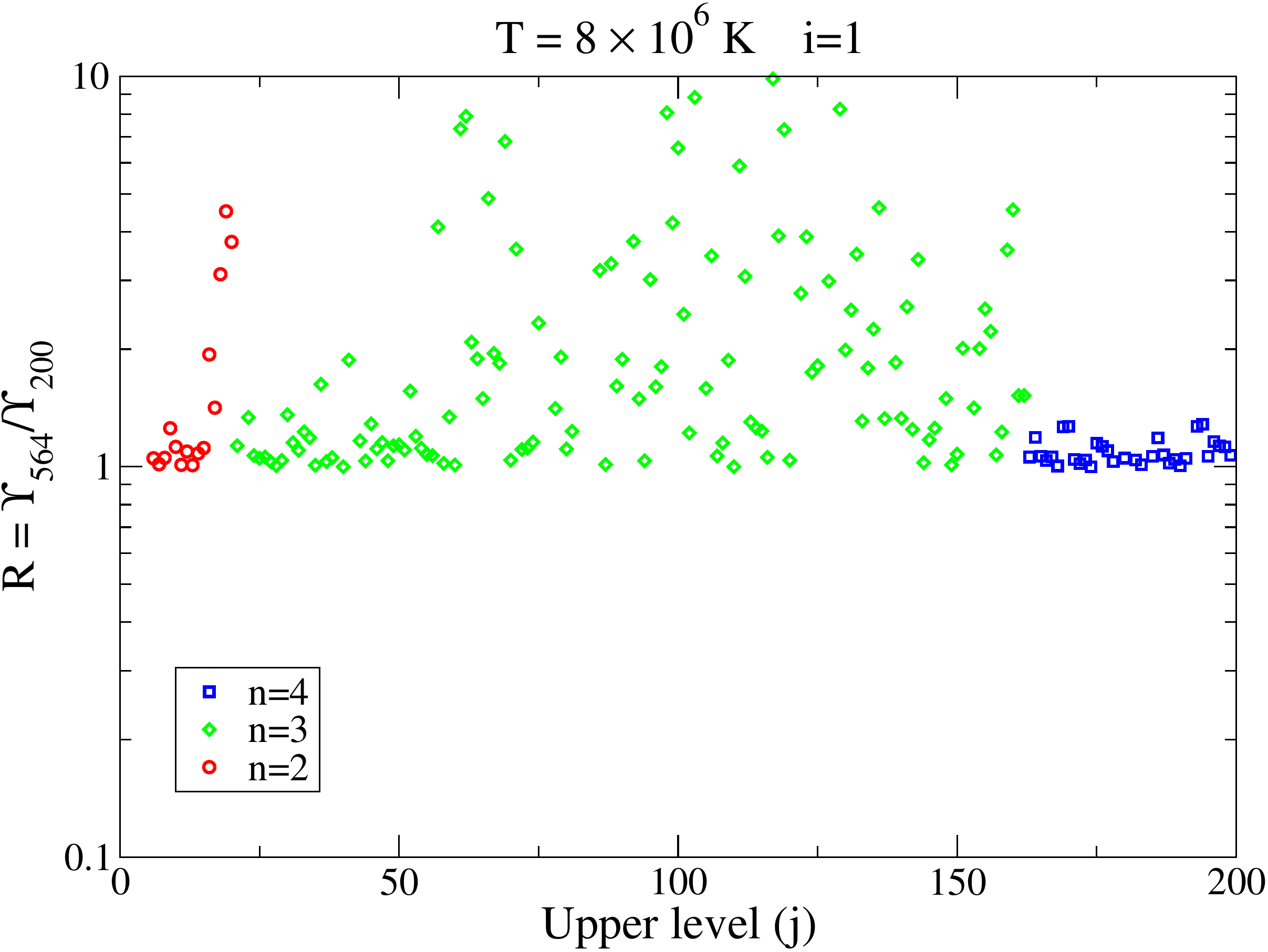}
   }\,
   \subfigure{
      \includegraphics[width=0.3\columnwidth]{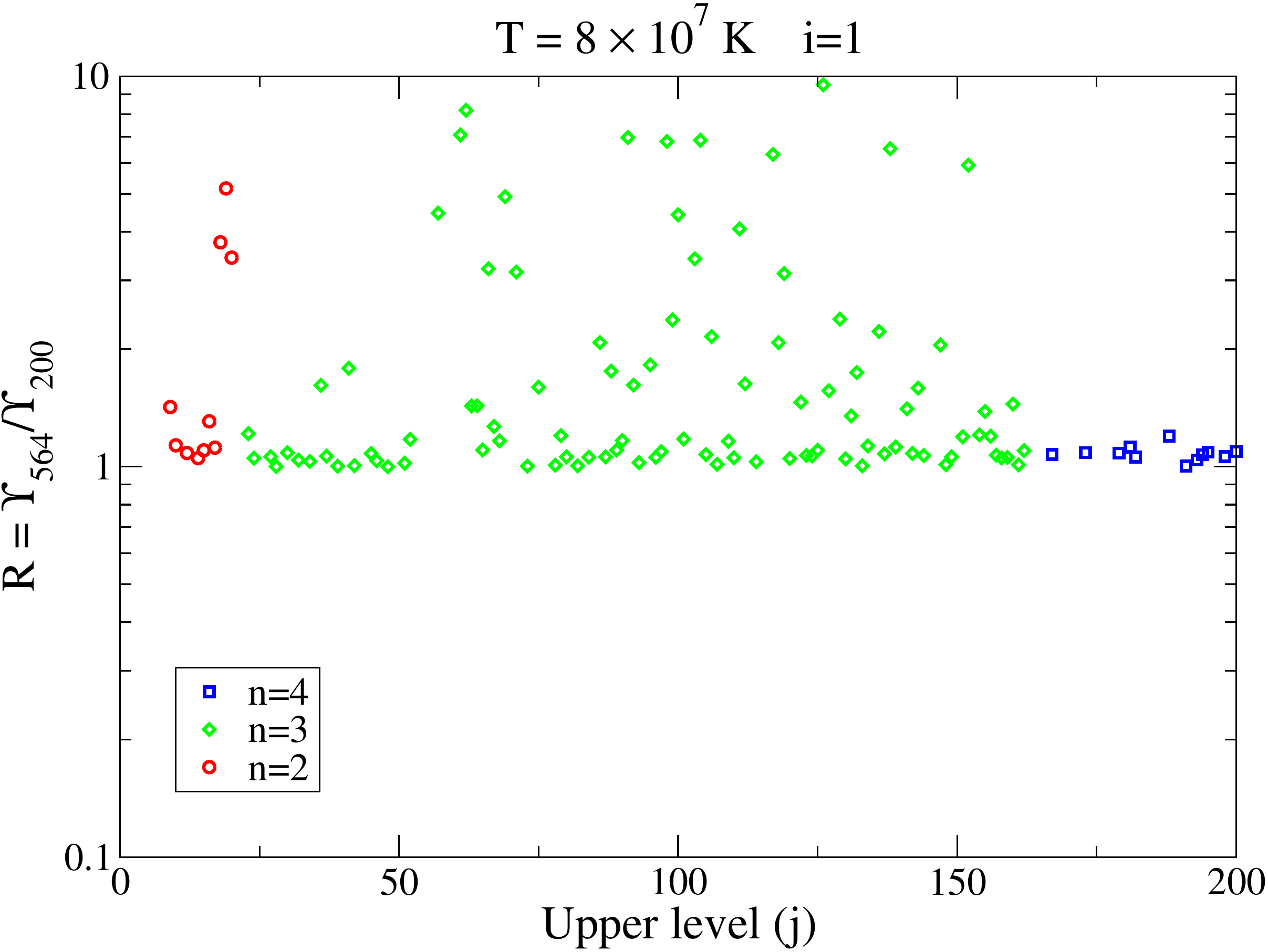}
   }
   \caption{Ratio of the $\Upsilon$ calculated with the two CC 
      expansions (present and  \cite{badnell2001a} )
      versus the upper level index from the ground level for three 
      temperatures.}
   \label{fig:upsk}
\end{figure}

Fig. \ref{fig:upsk} shows the same comparison, this time restricting just
for the transitions from the ground level.
It is these transitions which  \cite{badnell2001a} argued could be 
calculated with the reduced CC expansion.
Even using this restricted set of transitions the differences can be large,
up to a factor 10. It is notable that it is transitions involving $n=3$
which are affected most.
The largest differences lie in transitions with double electron jumps or
forbidden ones.
These differences can not be attributed to atomic structure (configuration 
mixing) as both calculations used exactly the same one.
The differences lie in the completeness of the close-coupling expansion.
The differences are also smaller as the temperature increases.
This is expected, as both sets of $\Omega$ tend to the same infinite energy limits.
Such a comparison for the astrophysically-relevant transitions from the other 
fine-structure levels of the ground term looks very similar.

\begin{figure}
\centering
   \subfigure{
      \includegraphics[width=0.3\columnwidth]{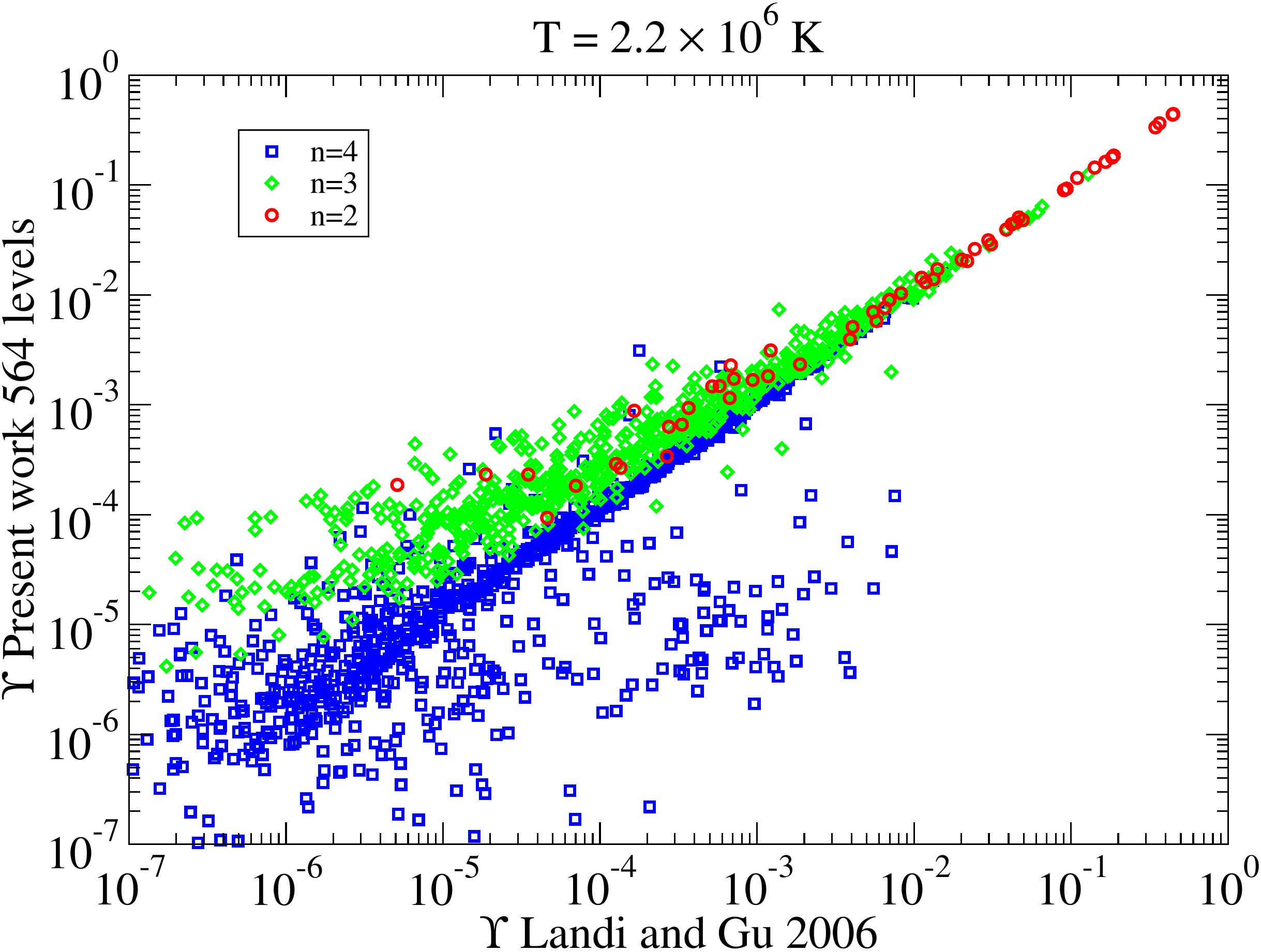}
   }\,
   \subfigure{
      \includegraphics[width=0.3\columnwidth]{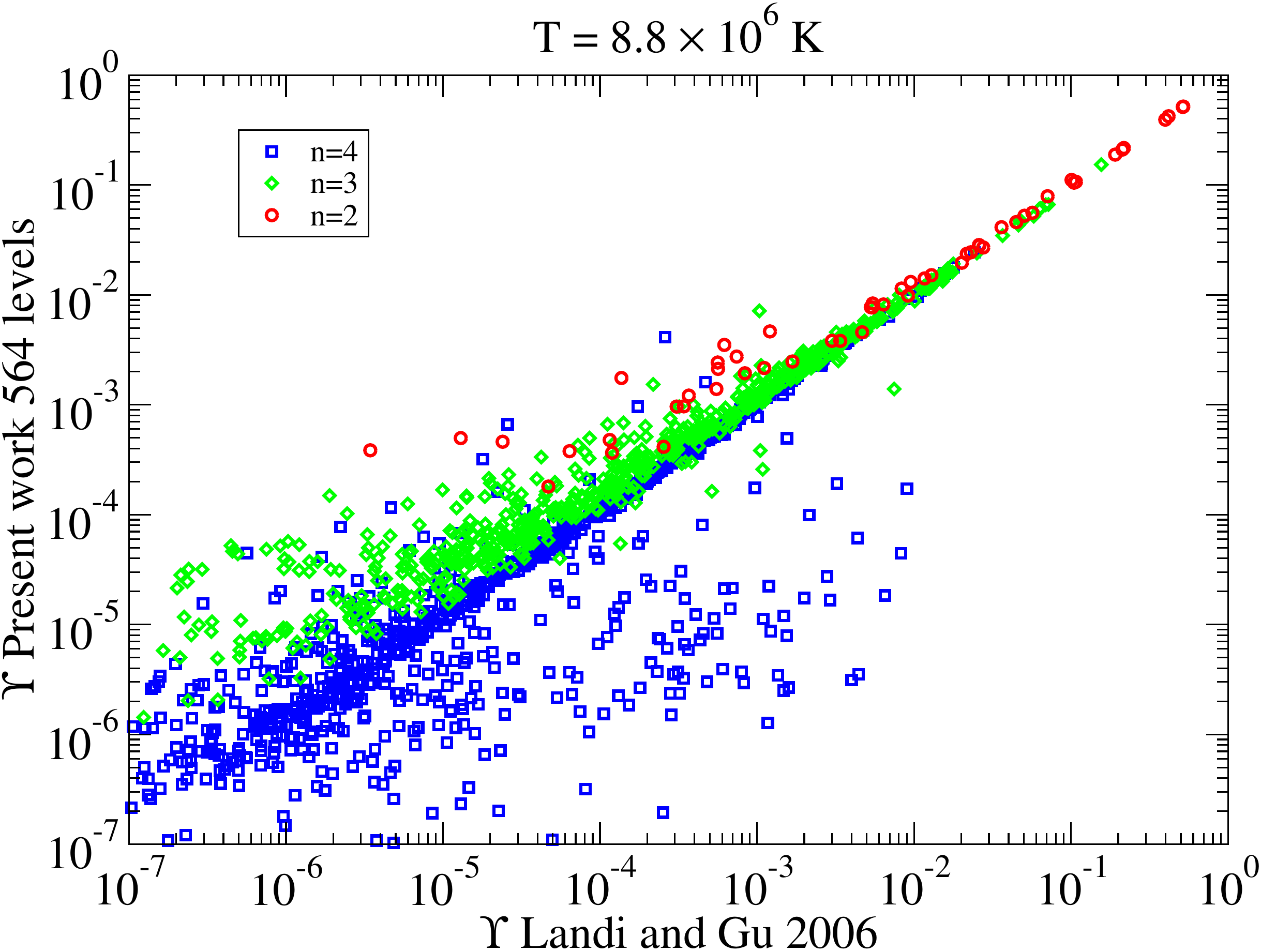}
   }\,
   \subfigure{
      \includegraphics[width=0.3\columnwidth]{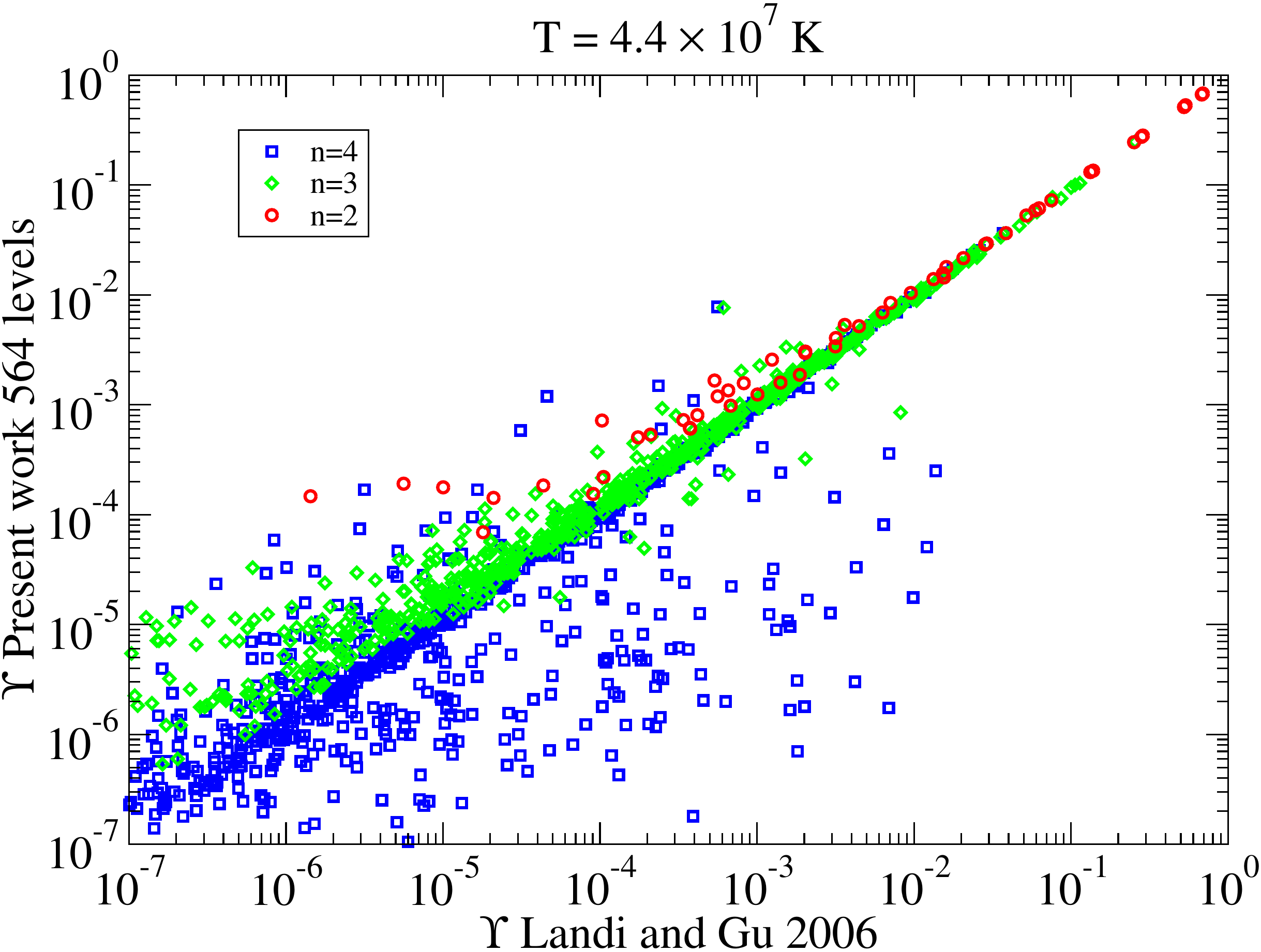}
   } \\
   \subfigure{
      \includegraphics[width=0.3\columnwidth]{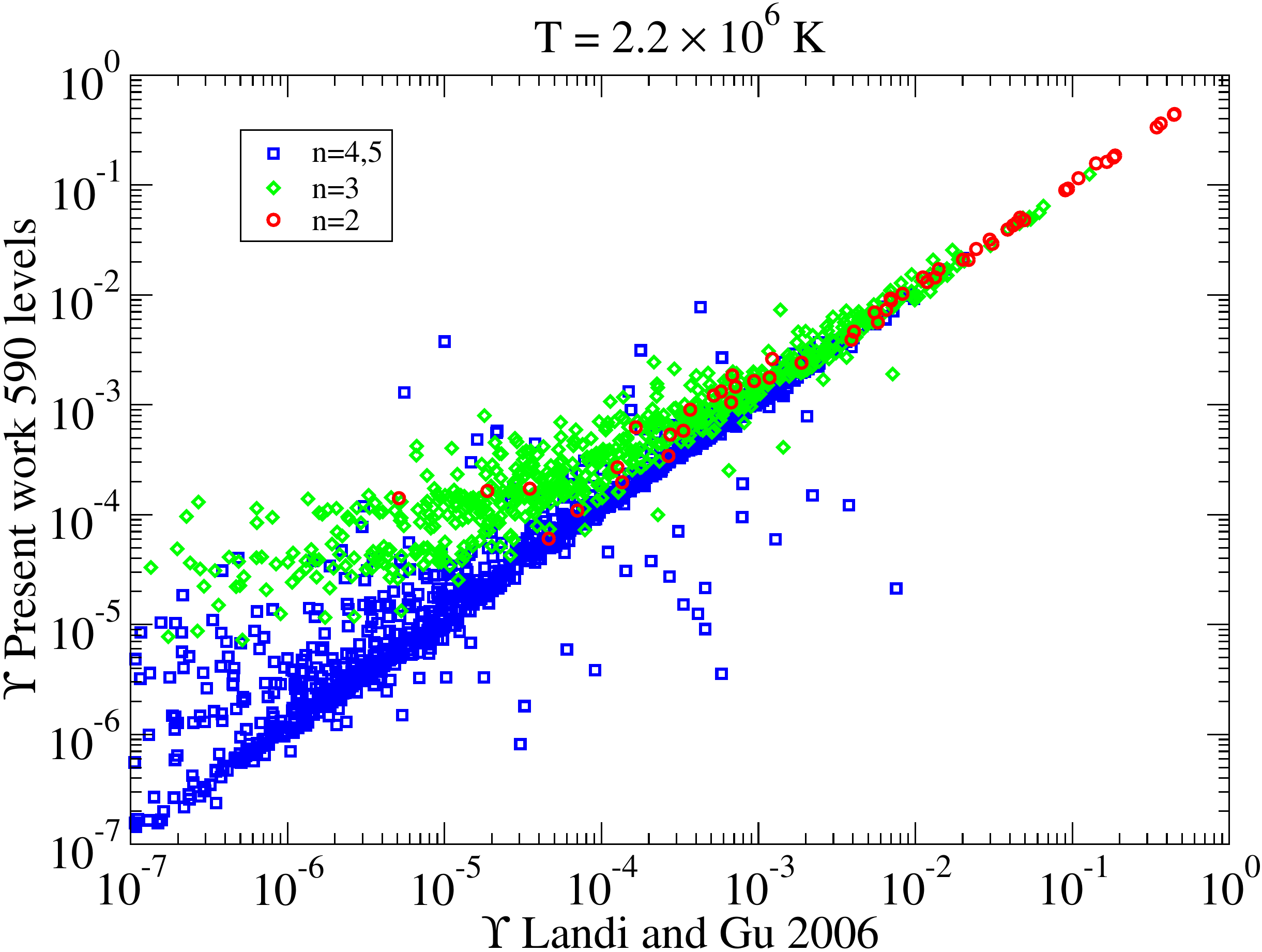}
   }\,
   \subfigure{
      \includegraphics[width=0.3\columnwidth]{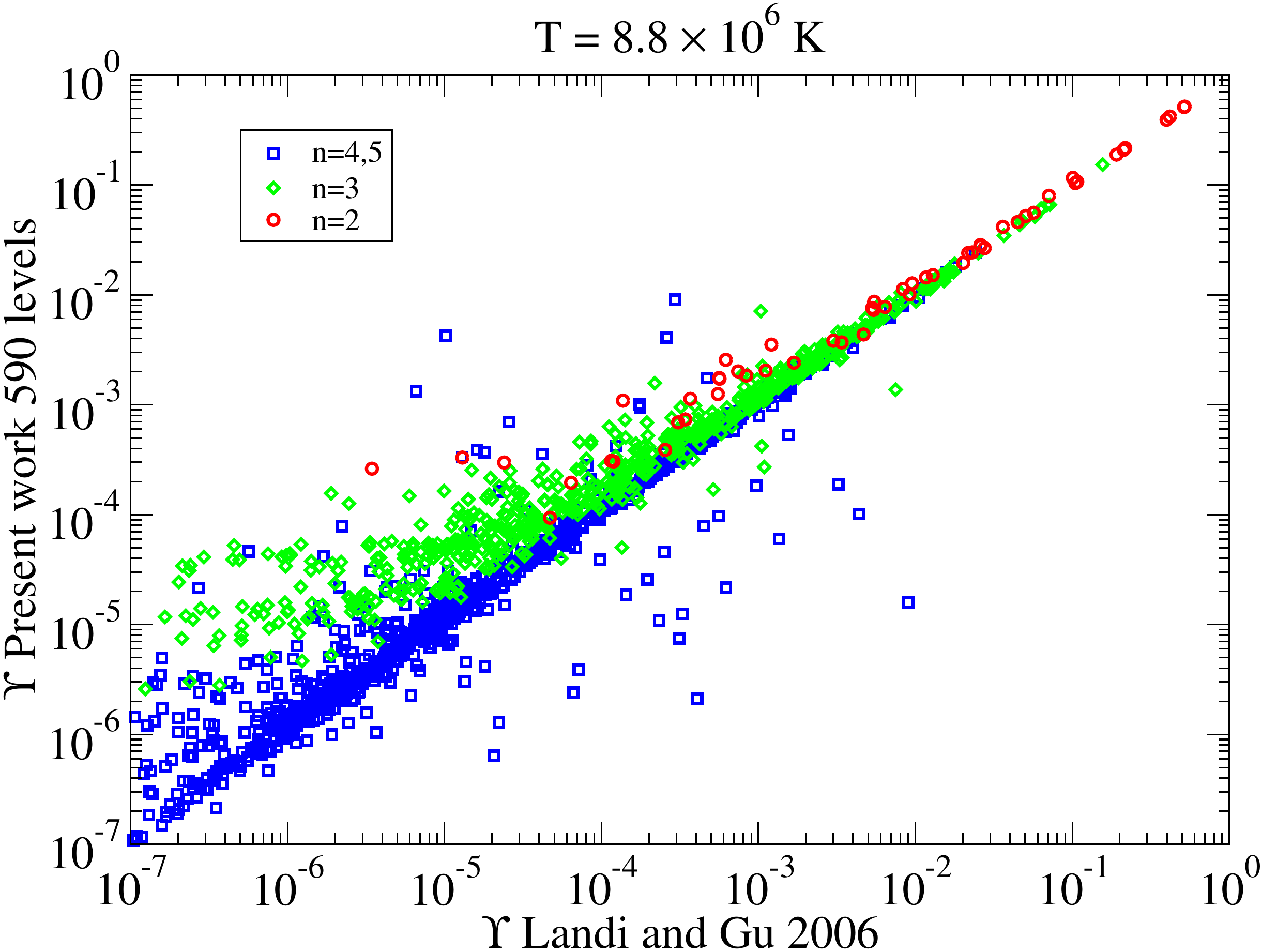}
   }\,
   \subfigure{
      \includegraphics[width=0.3\columnwidth]{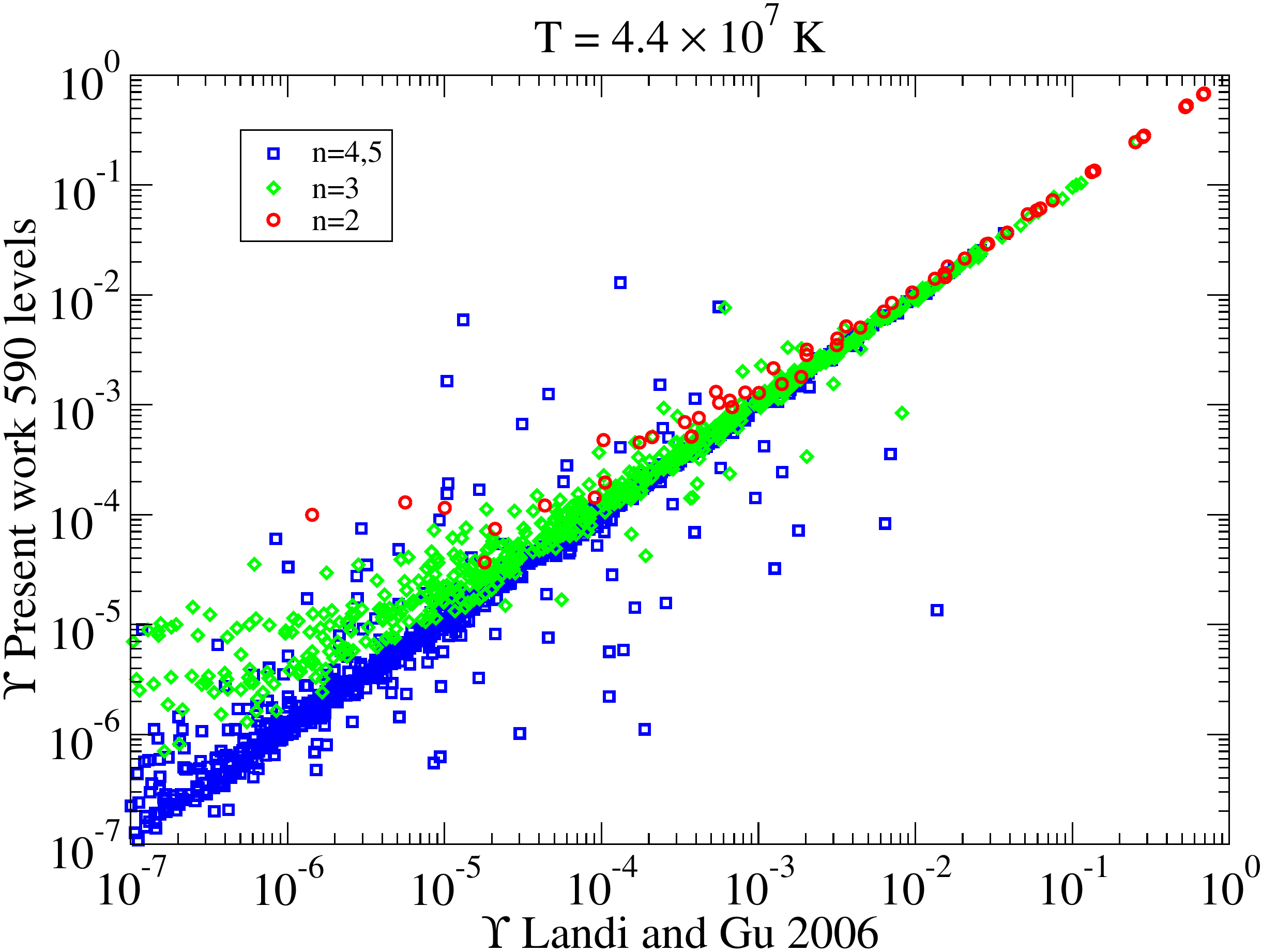}
   }
   \caption{Comparison for the $\Upsilon$ calculated in present work
      with the ones of Landi and Gu \cite{landi2006b}
      from the three lowest levels for three temperatures
      and the two CI/CC expansions.
      Above diagrams, comparison with the 564-level $R$-matrix calculation;
      below diagrams, comparison with the 590-level $R$-matrix calculation.}
   \label{fig:upslfmlan}
\end{figure}

In fig.~\ref{fig:upslfmlan} we compare our $R$-matrix results for both CI/CC 
expansions with the 590-level DW ones of Landi and Gu \cite{landi2006b}.
The dispersion in the diagrams comparing the DW with the 564-level $R$-matrix
calculation is much larger than that comparing with the 590-level $R$-matrix calculation.
In addition, in the comparison with the 590-level $R$-matrix calculation there are
very few points below the diagonal.
The main differences between the 590-level calculations can be attributed
to the additional resonance enhancement of the $R$-matrix calculation, while those
with the 564-level one for levels with $n=4$ are mainly related
to atomic structure --- compare with fig.~\ref{fig:gffe20k}.
The additional 26 extra levels in the CI / CC expansions improves
considerably the agreement of the collision data, as it did for the
atomic structure, for the levels  $\mathrm{2s^2\,2p}\,4l$ with $l=0-2$.

\begin{figure}
\centering
   \subfigure{
      \includegraphics[width=0.3\columnwidth]{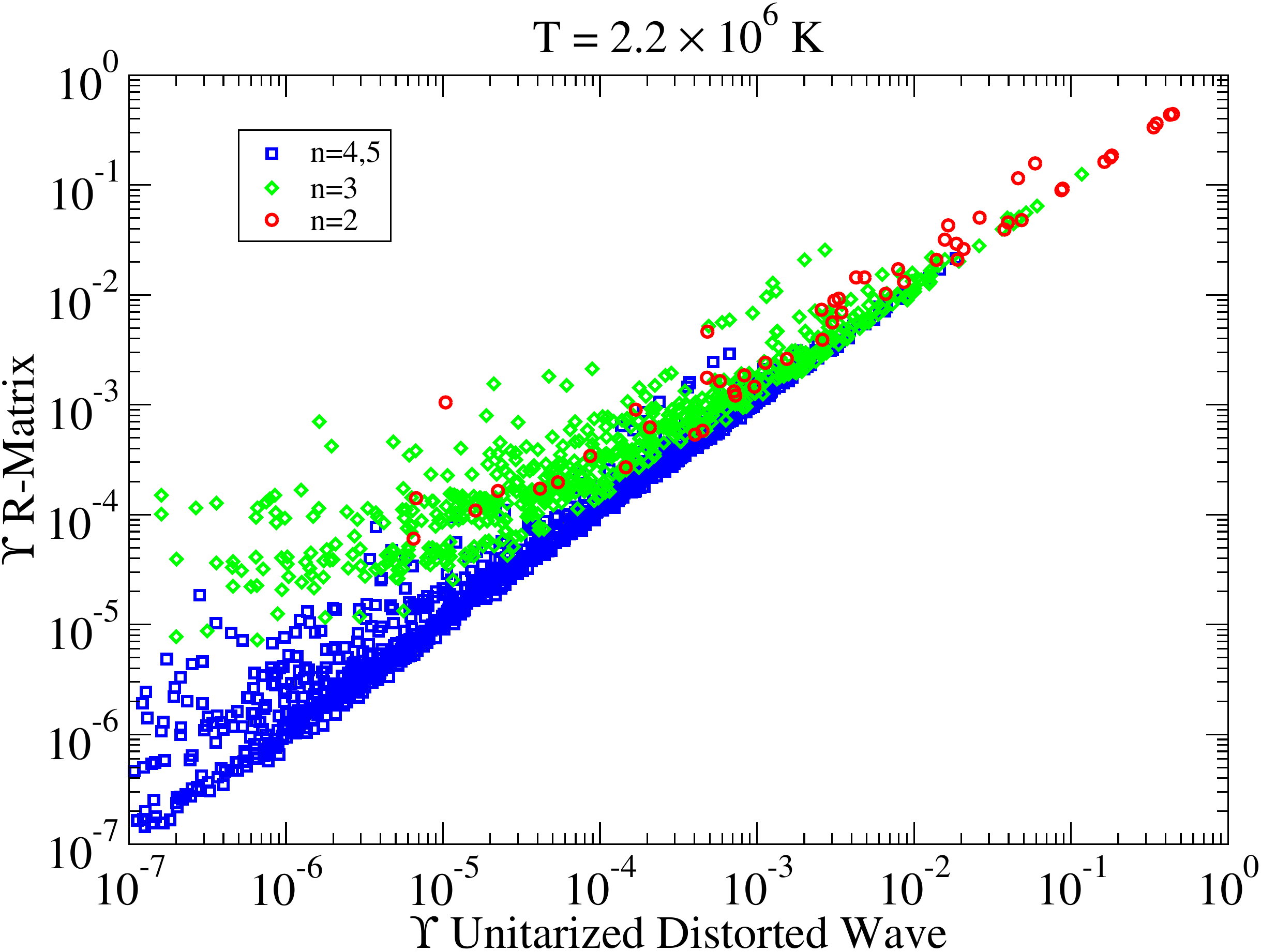}
   }\,
   \subfigure{
      \includegraphics[width=0.3\columnwidth]{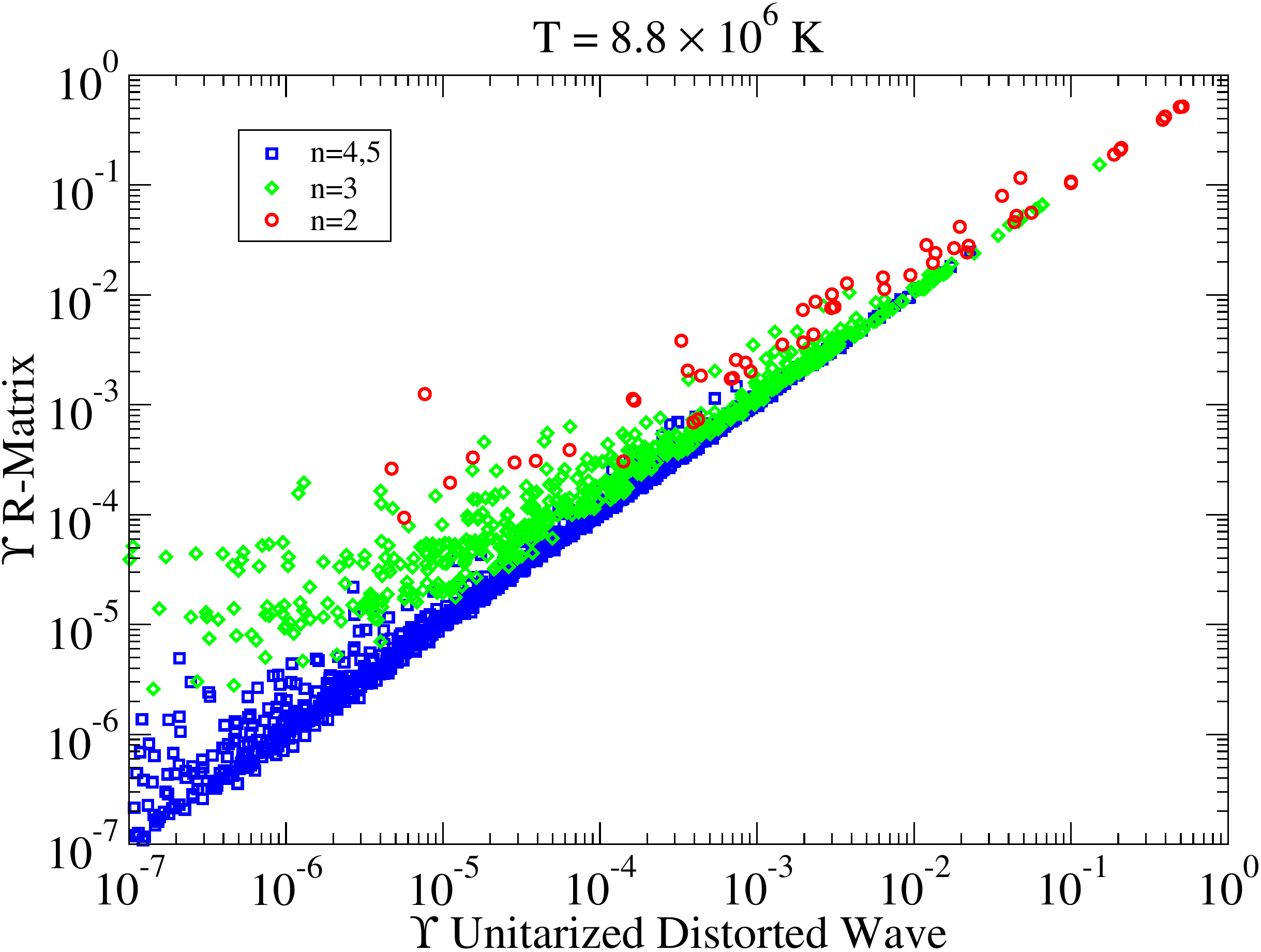}
   }\,
   \subfigure{
      \includegraphics[width=0.3\columnwidth]{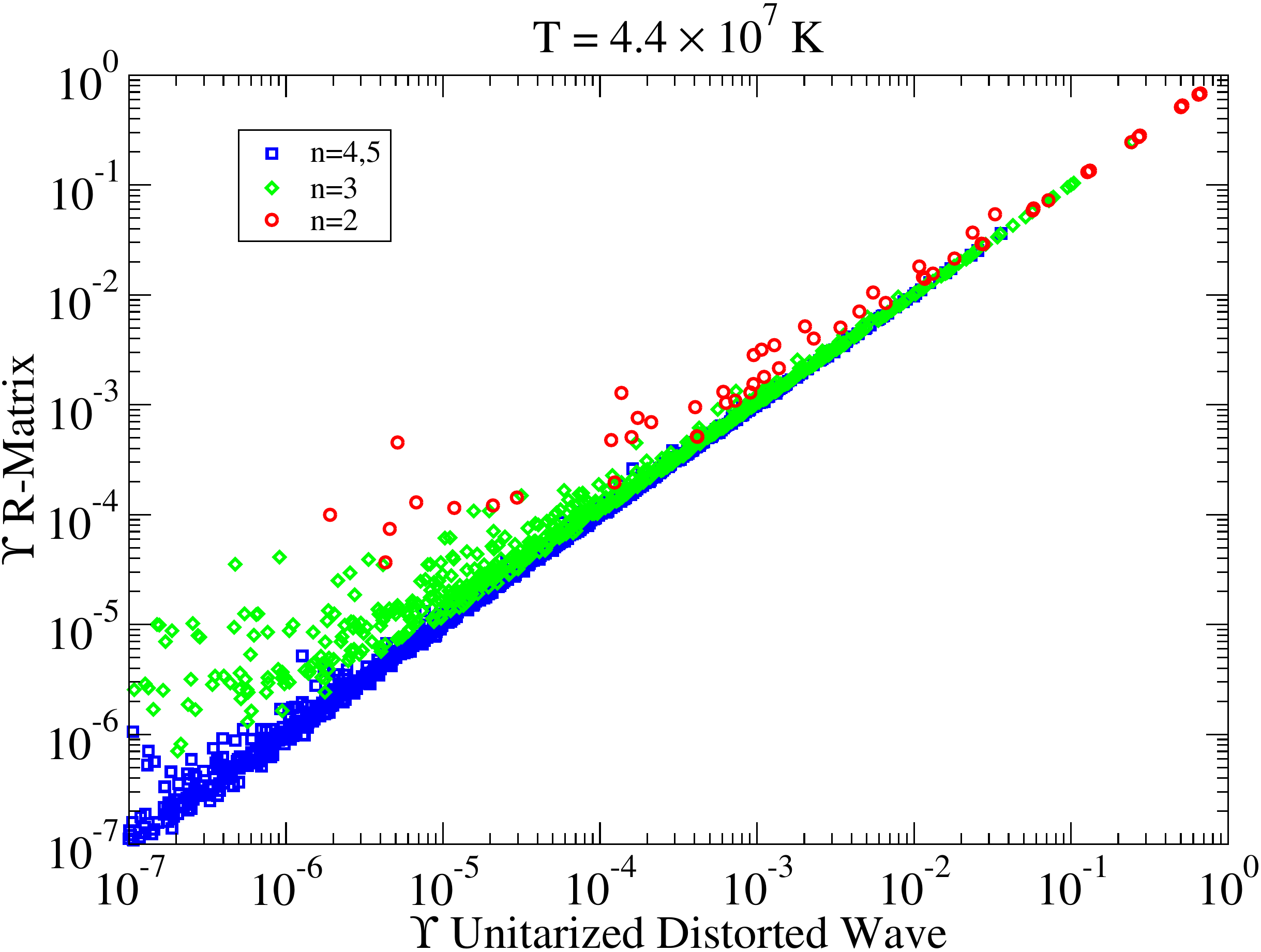}
   }
   \caption{Comparison for the $\Upsilon$ calculated in present work
      with the $R$-matrix and UDW methods with the same atomic structure of 
      the target (590 levels) from the three lowest levels for three 
      temperatures.}
   \label{fig:upsrmudw}
\end{figure}

The differences between the DW+resonances results of Landi and Gu \cite{landi2006b}
and the 590-level $R$-matrix ones are due to differences in atomic structure and
the difference between a perturbative and close-coupling treatment of resonances.
Thus, we have performed a non-resonant unitarized DW calculation using the same
atomic structure as the 590-level $R$-matrix one.
Fig.~\ref{fig:upsrmudw} shows the comparison between both calculations
$R$-matrix and UDW for the transitions tabulated in \cite{landi2006b}.
Now the only differences are due solely to the resonances, and strong coupling in general.
Qualitatively, the dispersion in fig.~\ref{fig:upsrmudw} is comparable to that in 
fig.~\ref{fig:upslfmlan} and, indeed, quantitatively (see table \ref{tab:denslan}) 
the level of disagreement between the two Distorted Wave calculations and 
$R$-matrix is very similar, despite one including resonances and the other
not. We note an earlier small study on Mg-like ions \cite{badnell1993}, which 
compared $R$-matrix with DW-plus-resonances utilizing identical atomic structure, found
significantly stronger resonance contributions from $R$-matrix due to interacting
resonances, i.e. a breakdown of the isolated resonance approximation used
by the perturbative DW approach.

\begin{table}
\begin{minipage}{\columnwidth}
   \caption{\label{tab:denslan} Number of transitions in Figs 
      \ref{fig:upslfmlan} and \ref{fig:upsrmudw} which 
      differ by more than a certain relative error
      $\delta=|\Upsilon-\Upsilon_{\mathrm{RM}}|/\Upsilon_{\mathrm{RM}}$ 
      as a percentage.
      $T=8.8 \times 10^6 \kelvin$.}
\begin{center}
\begin{tabular}{rrr}
   \hline 
     Rel. error ($\%$) & RM vs DW+res \cite{landi2006b} & RM vs UDW \\
   \hline
     10  & 1297  & 1127  \\
     20  & 1019  &  815  \\
     50  &  691  &  564  \\
    100  &  490  &  387  \\
    200  &  354  &  254  \\
    500  &  206  &  127  \\
   1000  &  124  &   71  \\
   \hline
   Total & 1764  & 1764  \\
   \hline
\end{tabular}                      
\end{center}
\end{minipage}
\end{table}

\begin{figure}
\centering
   \subfigure{
      \includegraphics[width=0.3\columnwidth]{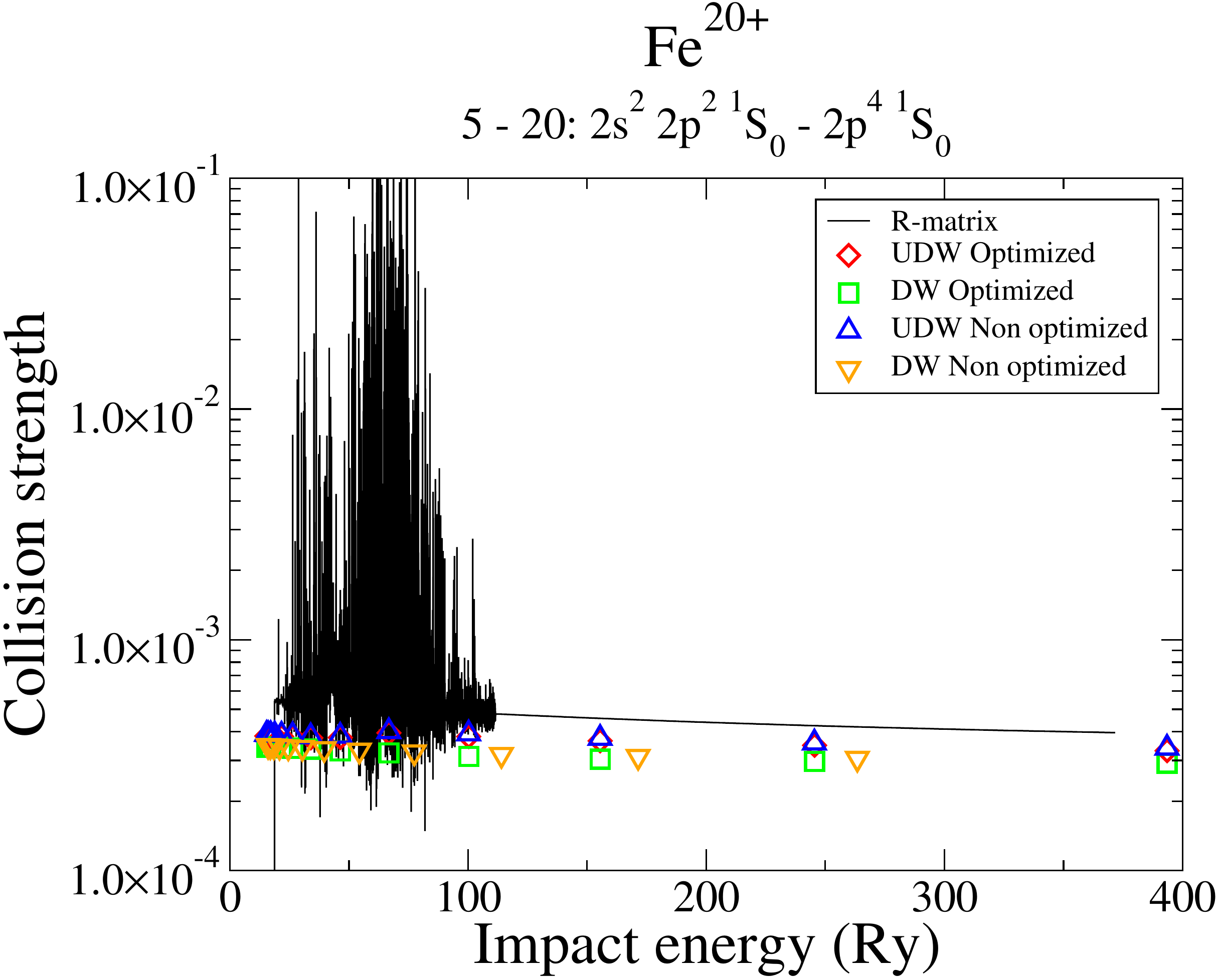}
   }\,
   \subfigure{
      \includegraphics[width=0.3\columnwidth]{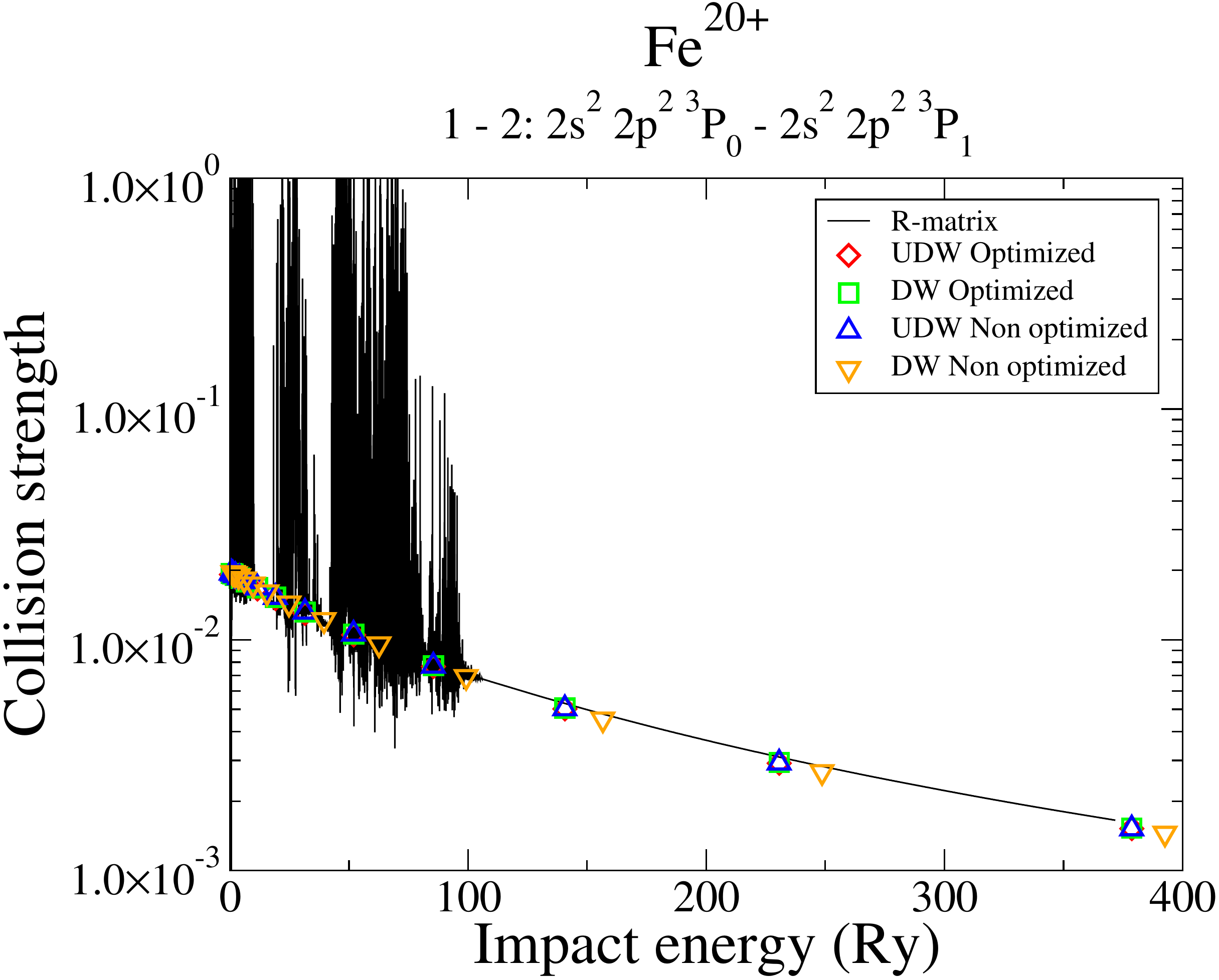}
   }
   \subfigure{
      \includegraphics[width=0.3\columnwidth]{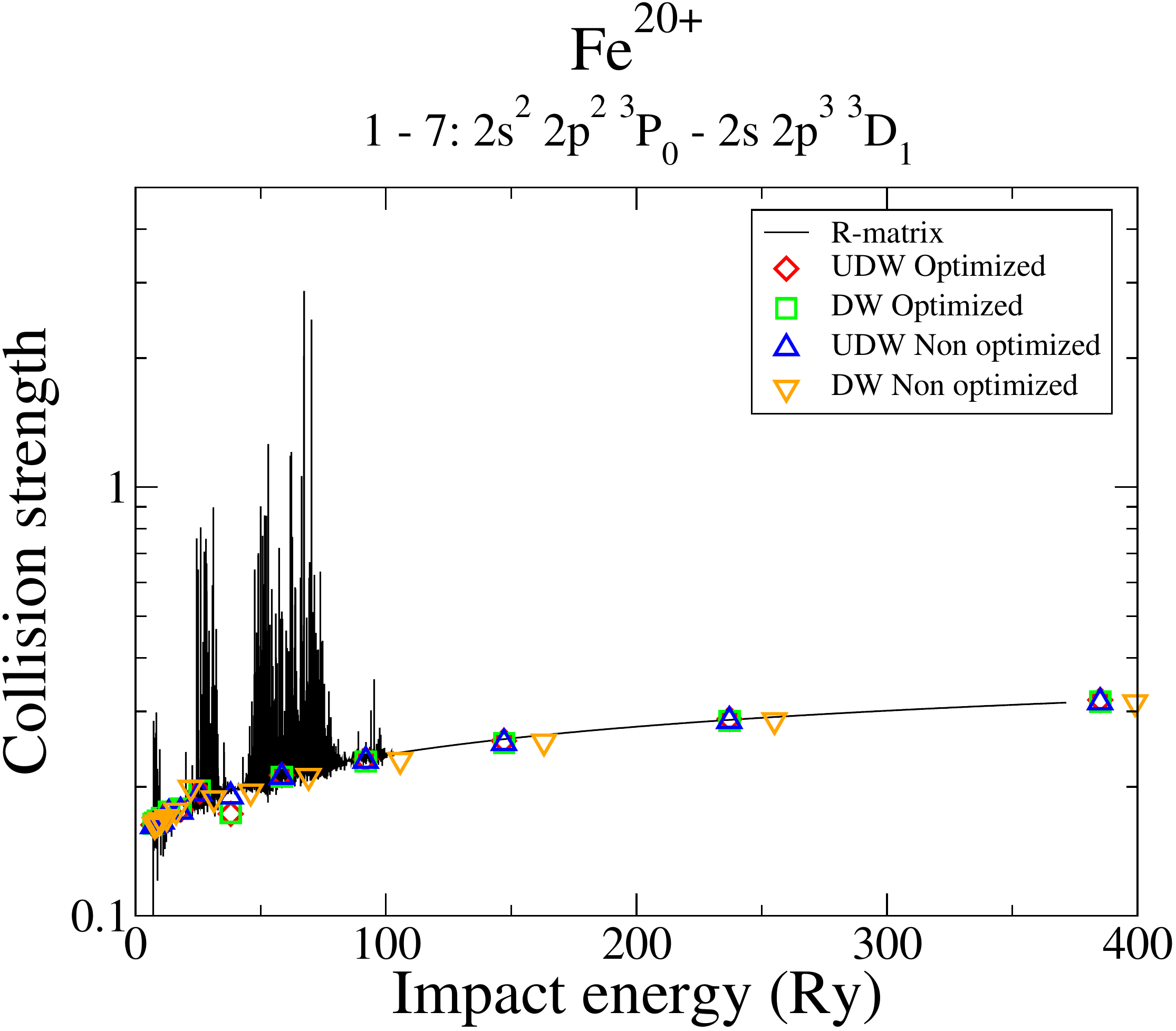}
   }
   \caption{Comparison of the $\Omega$ calculated with different optimized
      atomic structures and unitarized and non-unitarized formalism.
      564-level CI / CC expansions.
      Full line: $R$-matrix;
      $\diamond$: AS-UDW optimized atomic structure;
      $\square$: AS-DW optimized atomic structure;
      $\vartriangle$: AS-UDW simplified atomic structure;
      $\triangledown$: AS-DW simplified atomic structure;
      Colour online.}
   \label{fig:fe20omg}
\end{figure}

Fig. \ref{fig:fe20omg} shows a comparison between several DW calculations.
We show some selected transitions, 
the $5-20$: $\mathrm{2s^2 2p^2\,^1S_0 - 2p^4\,^1S_0}$ one
is an optically forbidden $J-J'=0-0$ one dominated by coupling.
It is the kind of transition which is sensitive to the unitarization of the
DW method \cite{fernandez-menchero2015a}.
We also compare two atomic structures, the optimized one, as it is explained
in section \ref{sec:structure}, and a simplified one, which has the same set
of configurations, but all the scaling parameters $\lambda_{nl}$ have been
fixed to unity.
The calculation of non-unitarized DW with the non-optimized atomic structure
is the one available in the OPEN-ADAS database since 2012.
DW calculations for both atomic structures agree to $\sim 1\%$.
The UDW underestimates the collision strength compared to the $R$-matrix 
one by $\sim 10\%$.
The non-unitarized DW calculation gives rise to a larger underestimate, of
$20\%$.
The effects of the scaling parameter is small in the final results, as
should perhaps be expected for such a relatively simple highly charged ion.
Thus, the DW data archived in OPEN-ADAS are valid for plasma 
modelling when transitions are not strongly resonance enhanced.

\begin{figure}
\centering
   \subfigure{
      \includegraphics[width=0.3\columnwidth]{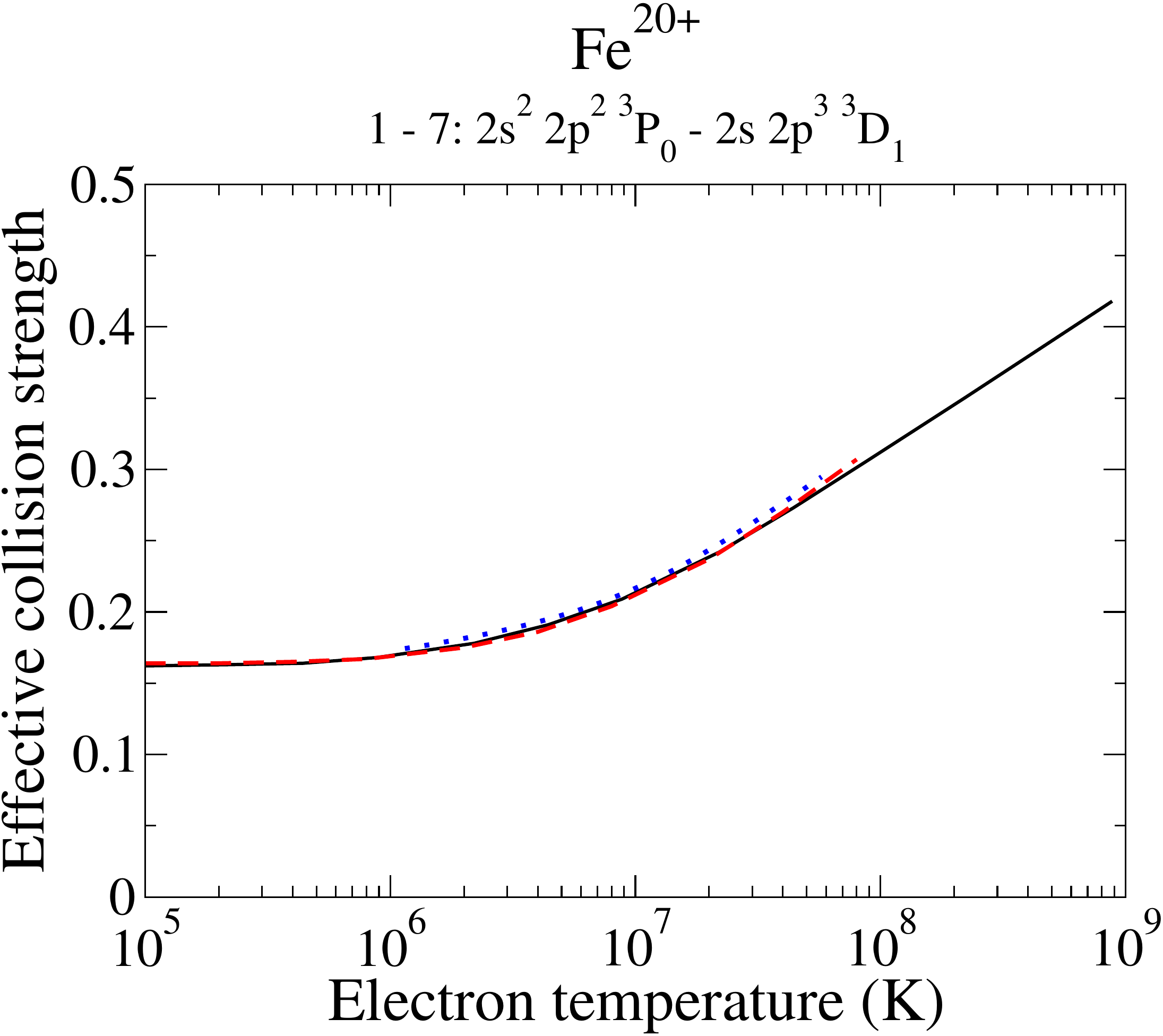}
   }\,
   \subfigure{
      \includegraphics[width=0.3\columnwidth]{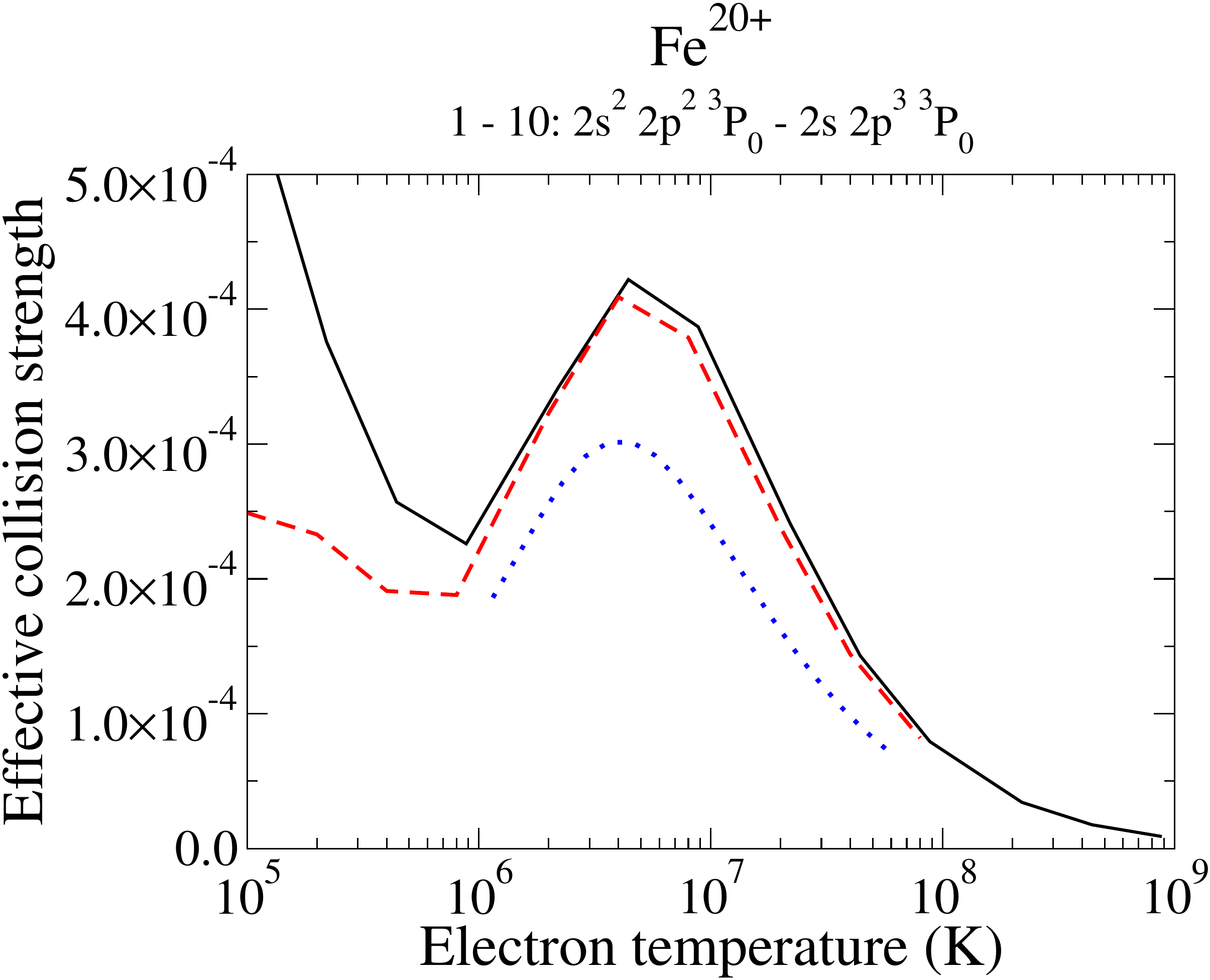}
   }\,
   \subfigure{
      \includegraphics[width=0.3\columnwidth]{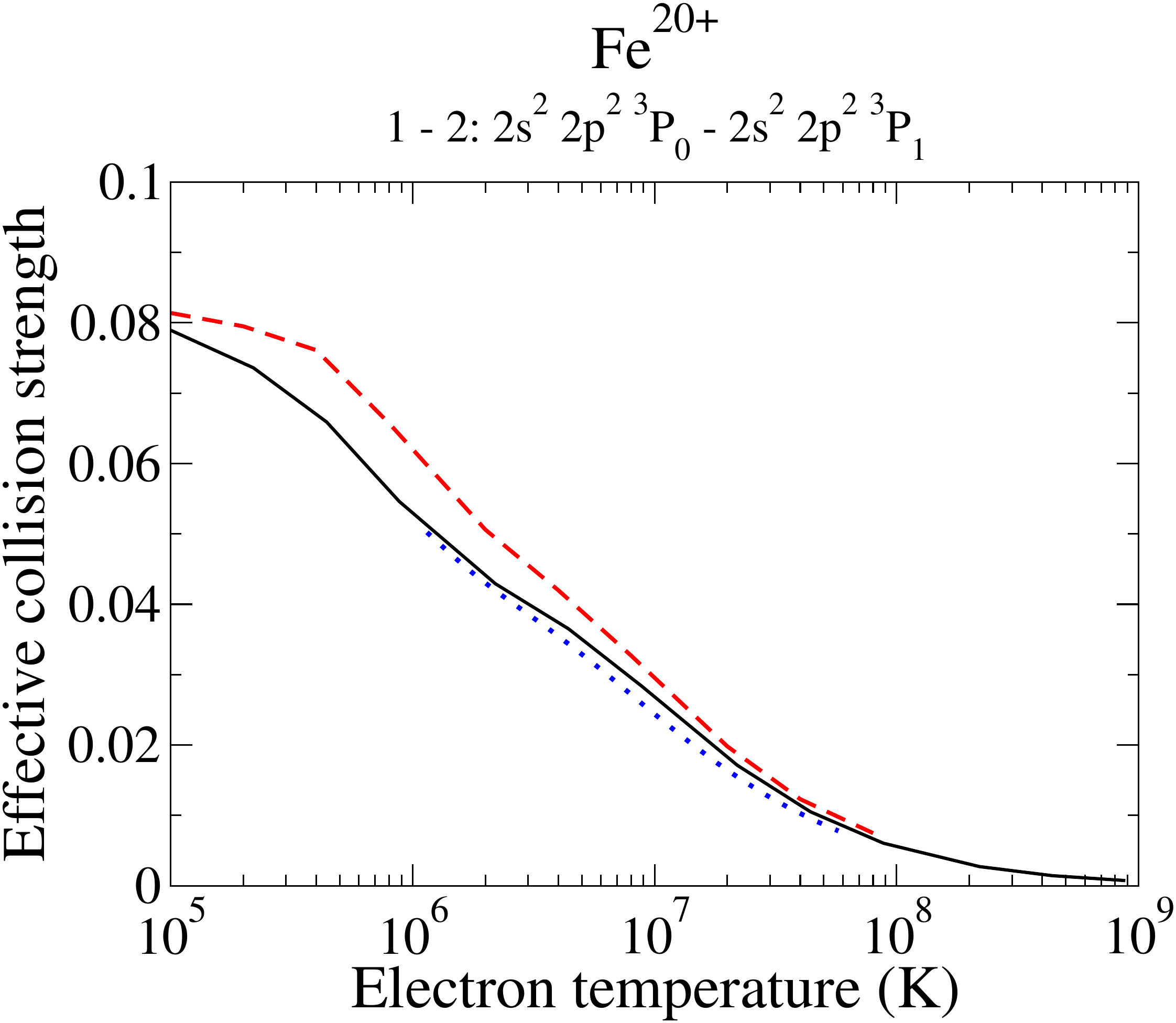}
   }
   \caption{Electron-impact excitation effective collision strengths versus 
      the electron temperature for some selected transitions within 
      the $n=2$ complex.
      Full line: present work; 
      dashed line: previous work \cite{badnell2001a};
      dotted line: work \cite{landi2006b}.
      Colour online.}
   \label{fig:ups}
\end{figure}

As a sample of the best results obtained, the ones of the 590-level 
$R$-matrix calculation, fig. \ref{fig:ups} shows the calculated effective 
collision strength compared with the previous works tabulated in CHIANTI
database \cite{badnell2001a,landi2006b} for some selected transitions.
We show the intense electric dipole transition $1-7$,
the forbidden M1 one which decays to the $1354.1\,\AA$ line $1-2$,
the one-photon forbidden $J-J'=0-0$ one $1-10$.
The rest of the values of the effective collision strengths for all 
of the $173\,755$  inelastic transitions can be found online.

For the electric dipole transition $1-7$, both CC expansions of the present
work and \cite{badnell2001a} lead to the same results.
The difference in the size of the CC expansions affects the resonance
region, and for intense dipole transitions it is a small contribution.
There is a difference between the $R$-matrix and the DW calculations for
the optically forbidden transition $1-10$ 
$\mathrm{2s^2\,2p^2\,^3P_0 - 2s\,2p^3\,^3P_0}$.
This is due to the resonance contribution.
Landi and Gu \cite{landi2006b} used the the independent-processes
and isolated-resonance approximations to include only resonances attached 
to the $n=2$ levels and some of the $n=3$, as discussed earlier.
The DW and $R$-matrix calculations lead to the same results for dipole transitions.
For the forbidden transition $1-10$ we appreciate a resonance enhancement
with respect to the old calculation \cite{badnell2001a} at low temperature.

\section{Conclusions}
\label{sec:conclusions}

We have calculated a complete data set for the electron-impact excitation of 
the $\mathrm{C}$-like ion $\mathrm{Fe}^{20+}$ using the ICFT $R$-matrix method.
We have checked the effects in the final results of including some relevant
configurations of $n=5$ in the CI expansion and the truncation of the 
CC expansion of the $R$-matrix calculation with respect to the CI expansion.
Including just 26 $n=5$ levels in the CI / CC expansions does not increase 
substantially the computing resources required, but it improves significantly
the convergence of the configuration interaction expansion of the target, for the 
levels  $\mathrm{2s^2\,2p}\,4l$ with $l=0-2$.
The truncated CC expansion of \cite{badnell2001a} underestimates substantially
the results between excited states, but transitions from the ground 
configuration less so, mainly for the weak transitions and low temperatures.
This is due to the coupling with states in the larger expansion which are
not present in the smaller one.

The DW+resonances data of Landi and Gu \cite{landi2006b} were to-date the most 
extensive and accurate calculation for the electron-impact excitation effective 
collision strengths of $\mathrm{Fe}^{20+}$ for transitions from the three
lowest levels.
The present calculation uses an atomic structure of the same quality as that
of \cite{landi2006b} and calculates the whole transition matrix with an
$R$-matrix method. This full treatment of coupling and resonances gives
rise to significant differences (increases, generally) compared to the
results of Landi and Gu \cite{landi2006b} while differences due to
the remaining differences in atomic structure are likely (much) smaller.

We have also compared our DW results for an optimized atomic structure with 
an earlier non-optimized one.
Results for both atomic structures are generally in good agreement.
The optimization of the atomic structure does not affect significantly in
the final results for such a highly-charged ion. In this sense, the
DW data for $\mathrm{Fe}^{20+}$ present in OPEN-ADAS database since 2012 can
be used for plasma modelling. However, we find significant differences
between this DW data and the present $R$-matrix data due to the omission
of resonances, and coupling more generally.

Even in Fe$^{20+}$, it is necessary to ensure that the CI and CC expansions 
are converged sufficiently. Thus,
in the future, we plan apply this same $R$-matrix approach to 
$\mathrm{C}$-like $\mathrm{Ni}^{22+}$ for astrophysical modelling applications.

\section*{Acknowledgments}

The present work was funded by STFC (UK) through the 
University of Strathclyde UK APAP Network grant
ST/J000892/1,
the University of Cambridge DAMTP grant and
the European Framework 7 ADAS-EU support action.

\section*{References}

\bibliographystyle{iopart-num}
\bibliography{references}

\providecommand{\newblock}{}
\begin{thebibliography}{10}
\expandafter\ifx\csname url\endcsname\relax
  \def\url#1{{\tt #1}}\fi
\expandafter\ifx\csname urlprefix\endcsname\relax\def\urlprefix{URL }\fi
\providecommand{\eprint}[2][]{\url{#2}}

\bibitem{depontieu2014}
{De Pontieu} B, {Title} A~M, {Lemen} J~R, {Kushner} G~D, {Akin} D~J, {Allard}
  B, {Berger} T, {Boerner} P, {Cheung} M, {Chou} C, {Drake} J~F, {Duncan} D~W,
  {Freeland} S, {Heyman} G~F, {Hoffman} C, {Hurlburt} N~E, {Lindgren} R~W,
  {Mathur} D, {Rehse} R, {Sabolish} D, {Seguin} R, {Schrijver} C~J, {Tarbell}
  T~D, {W{\"u}lser} J~P, {Wolfson} C~J, {Yanari} C, {Mudge} J, {Nguyen-Phuc} N,
  {Timmons} R, {van Bezooijen} R, {Weingrod} I, {Brookner} R, {Butcher} G,
  {Dougherty} B, {Eder} J, {Knagenhjelm} V, {Larsen} S, {Mansir} D, {Phan} L,
  {Boyle} P, {Cheimets} P~N, {DeLuca} E~E, {Golub} L, {Gates} R, {Hertz} E,
  {McKillop} S, {Park} S, {Perry} T, {Podgorski} W~A, {Reeves} K, {Saar} S,
  {Testa} P, {Tian} H, {Weber} M, {Dunn} C, {Eccles} S, {Jaeggli} S~A,
  {Kankelborg} C~C, {Mashburn} K, {Pust} N, {Springer} L, {Carvalho} R,
  {Kleint} L, {Marmie} J, {Mazmanian} E, {Pereira} T~M~D, {Sawyer} S, {Strong}
  J, {Worden} S~P, {Carlsson} M, {Hansteen} V~H, {Leenaarts} J, {Wiesmann} M,
  {Aloise} J, {Chu} K~C, {Bush} R~I, {Scherrer} P~H, {Brekke} P,
  {Martinez-Sykora} J, {Lites} B~W, {McIntosh} S~W, {Uitenbroek} H, {Okamoto}
  T~J, {Gummin} M~A, {Auker} G, {Jerram} P, {Pool} P and {Waltham} N 2014 {\em
  Solar Physics\/} {\bf 289} 2733--2779 (\textit{Preprint} \eprint{1401.2491})

\bibitem{young2015}
{Young} P~R, {Tian} H and {Jaeggli} S 2015 {\em Astrophys. J.\/} {\bf 799} 218
  (\textit{Preprint} \eprint{1409.8603})

\bibitem{polito2015}
{Polito} V, {Reeves} K~K, {Del Zanna} G, {Golub} L and {Mason} H~E 2015 {\em
  Astrophys. J.\/} {\bf 803} 84

\bibitem{mason1984}
{Mason} H~E, {Bhatia} A~K, {Neupert} W~M, {Swartz} M and {Kastner} S~O 1984
  {\em Solar Physics\/} {\bf 92} 199--216

\bibitem{delzanna2013}
{Del Zanna} G and {Woods} T~N 2013 {\em Astron. Astrophys.\/} {\bf 555} A59

\bibitem{odwyer2010}
{O'Dwyer} B, {Del Zanna} G, {Mason} H~E, {Weber} M~A and {Tripathi} D 2010 {\em
  Astron. Astrophys.\/} {\bf 521} A21

\bibitem{petkaki2012}
{Petkaki} P, {Del Zanna} G, {Mason} H~E and {Bradshaw} S 2012 {\em Astron.
  Astrophys.\/} {\bf 547} A25

\bibitem{hummer1993}
Hummer D~G, Berrington K~A, Eissner W, Pradhan A~K, Saraph H~E and Tully J~A
  1993 {\em Astron. Astrophys.\/} {\bf 279} 298--309

\bibitem{bhatia1987}
Bhatia A~K, Seely J~F and Feldman U 1987 {\em Atomic Data and Nuclear Data
  Tables\/} {\bf 36} 453

\bibitem{zhang1996}
Zhang H~L and Sampson D~H 1996 {\em Atomic Data and Nuclear Data Tables\/} {\bf
  63} 275 -- 314 ISSN 0092-640X/96

\bibitem{aggarwal1991}
Aggarwal K~M 1991 {\em Astrophys. J. Suppl. Ser.\/} {\bf 77} 677--696

\bibitem{aggarwal1999b}
Aggarwal K~M and Keenan F~P 1999 {\em J. Phys. B\/} {\bf 32} 3585

\bibitem{badnell2001a}
Badnell N~R and Griffin D~C 2001 {\em J. Phys. B\/} {\bf 34} 681--697

\bibitem{griffin1998}
Griffin D~C, Badnell N~R and Pindzola M~S 1998 {\em J. Phys. B\/} {\bf 31}
  3713--3727

\bibitem{landi2006b}
Landi E and Gu M~F 2006 {\em Astrophys. J.\/} {\bf 640} 1171--1179

\bibitem{fernandez-menchero2015b}
Fern\'andez-Menchero L, {Del~Zanna} G and Badnell N~R 2015 {\em Mon. Not. R.
  Astr. Soc.\/} {\bf 450} 4174--4183

\bibitem{badnell2011b}
Badnell N~R 2011 {\em Comput. Phys. Commun.\/} {\bf 182} 1528--1535

\bibitem{landi2013}
Landi E, Young P~R, Dere K~P, {Del~Zanna} G and Mason H~E 2013 {\em Astrophys.
  J.\/} {\bf 763} 86

\bibitem{fernandez-menchero2014a}
Fern\'andez-Menchero L, {Del~Zanna} G and Badnell N~R 2014 {\em Astron.
  Astrophys.\/} {\bf 566} A104

\bibitem{fernandez-menchero2014b}
Fern\'andez-Menchero L, {Del~Zanna} G and Badnell N~R 2014 {\em Astron.
  Astrophys.\/} {\bf 572} A115

\bibitem{witthoeft2007}
Witthoeft M~C, Whiteford A~D and Badnell N~R 2007 {\em J. Phys. B: At. Mol.
  Opt. Phys.\/} {\bf 40} 2969--2993

\bibitem{liang2010a}
Liang G~Y and Badnell N~R 2010 {\em Astron. Astrophys.\/} {\bf 518} A64

\bibitem{liang2011}
Liang G~Y and Badnell N~R 2011 {\em Astron. Astrophys.\/} {\bf 528} A69

\bibitem{liang2009a}
Liang G~Y, Whiteford A~D and Badnell N~R 2009 {\em Astron. Astrophys.\/} {\bf
  499} 943--954

\bibitem{eissner1974}
Eissner W~M, Jones M and H N 1974 {\em Comput. Phys Commun.\/} {\bf 8} 270

\bibitem{feldman2000a}
Feldman U, Curdt W, Landi E and Wilhelm K 2000 {\em The Astrophysical
  Journal\/} {\bf 544} 508
  \urlprefix\url{http://stacks.iop.org/0004-637X/544/i=1/a=508}

\bibitem{martin1999b}
Martin W~C, Fuhr J~R, Kelleher D~E, Musgrove A, Sugar J, Wiese W~L, Mohr P~J
  and Olsen K~J 1999 {\em NIST Physical Reference Data\/}
  \urlprefix\url{http://physics.nist.gov/asd}

\bibitem{brown2002}
Brown G~V, Beiersdorfer P, Liedahl D~A, Widmann K, Kahn S~M and Clothiaux E~J
  2002 {\em The Astrophysical Journal Supplement Series\/} {\bf 140} 589
  \urlprefix\url{http://stacks.iop.org/0067-0049/140/i=2/a=589}

\bibitem{fawcett1987}
Fawcett B~C, Jordan c, Lemen J and Phillips K 1987 {\em Monthly Notices of the
  Royal Astronomical Society\/} {\bf 225} 1013--1023

\bibitem{landi2005}
Landi E and Phillips K~J~H 2005 {\em The Astrophysical Journal Supplement
  Series\/} {\bf 160} 286
  \urlprefix\url{http://stacks.iop.org/0067-0049/160/i=1/a=286}

\bibitem{palmeri2003}
{P Palmeri}, {C Mendoza}, {T~R Kallman} and {M~A Bautista} 2003 {\em Astron.
  Astrophys.\/} {\bf 403} 1175--1184

\bibitem{berrington1995}
Berrington K~A, Eissner W~B and Norrington P~H 1995 {\em Comput. Phys.
  Commun.\/} {\bf 92} 290

\bibitem{badnell2001b}
Badnell N~R, Griffin D~C and Mitnik D~M 2001 {\em Journal of Physics B: Atomic,
  Molecular and Optical Physics\/} {\bf 34} 5071
  \urlprefix\url{http://stacks.iop.org/0953-4075/34/i=24/a=309}

\bibitem{burgess1974}
Burgess A 1974 {\em Journal of Physics B: Atomic and Molecular Physics\/} {\bf
  7} L364 \urlprefix\url{http://stacks.iop.org/0022-3700/7/i=12/a=003}

\bibitem{burgess1992}
Burgess A and Tully J~A 1992 {\em Astron. Astrophys.\/} {\bf 254} 436--453
  \urlprefix\url{http://adsabs.harvard.edu/abs/1992A\%26A...254..436B}

\bibitem{summers1994}
Summers H~P 1994 {\em ADAS manual\/} JET Joint Undertaking

\bibitem{badnell1993}
Badnell N~R, Girffin D~C, Gorczyca T~W and Pindzola M~S 1993 {\em Phys. Rev.
  A\/} {\bf 48} R2519

\bibitem{fernandez-menchero2015a}
Fern\'andez-Menchero L, {Del~Zanna} G and Badnell N~R 2015 {\em Astron.
  Astrophys.\/} {\bf 577} A95

\end{thebibliography}


\end{document}